\documentclass[11pt]{article}
\usepackage{authblk}
\usepackage[ansinew]{inputenc}
\usepackage{amsmath, amssymb, graphics, amsthm}
\usepackage{epsfig}
\usepackage{color}
\usepackage{fancyhdr}
\usepackage{stmaryrd}
\usepackage{amsfonts}
\usepackage{mathrsfs}
\usepackage{txfonts}
\usepackage{braket}
\usepackage{tensor}
\usepackage{comment}
\usepackage{dsfont}
\usepackage{euscript}

\usepackage[normalem]{ulem}

\newcommand{\abs}[1]{{\left|{#1}\right|}} 
\newcommand{\kt}{{\tilde{K}}} 
\newcommand{\R}{\mathcal {R}} 
\newcommand{\kin}{\mathrm{kin}} 
\newcommand{\hil}{\mathcal{H}} 
\newcommand{\Euc}{H^{E}} 
\newcommand{\grav}{\mathrm{gr}} 
\newcommand{\ints}{{\int_\Sigma}} 
\newcommand{\sca}{\mathrm{sc}} 
\newcommand{\Tr}{\mathrm{Tr}} 

\makeatletter
\@addtoreset{equation}{section}
\makeatother

\newcommand{\be}{\begin{equation}}
\newcommand{\ee}{\end{equation}}
\newcommand{\ba}{\begin{eqnarray}}
\newcommand{\ea}{\end{eqnarray}}
\def\nn{\nonumber}

\def\pb#1{\rlap{\lower1.5ex\hbox{$\longleftarrow$}}{#1}}
\def\dpb#1{\rlap{\lower1.5ex\hbox{$\Longleftarrow$}}{#1}}
\def\spb#1{\rlap{\lower1.0ex\hbox{$\leftarrow$}}{#1}}
\def\sdpb#1{\rlap{\lower1.0ex\hbox{$\Leftarrow$}}{#1}}

\title{{\sf Hamiltonian Theory: generalizations to higher dimensions, supersymmetry and modified gravity}}

\author[1]{\textsf{Norbert Bodendorfer\footnote{\texttt{norbert.bodendorfer@physik.uni-r.de}}}}
\author[2]{\textsf{Konstantin Eder\footnote{\texttt{konstantin.eder@gravity.fau.de}}}}
\author[3]{\textsf{Xiangdong Zhang\footnote{\texttt{scxdzhang@scut.edu.cn}}}}
\affil[1]{\textsf{Institute for Theoretical Physics, University of Regensburg,}
\protect\\ \textsf{93040 Regensburg, Germany}}
\affil[2]{\textsf{Institute for Quantum Gravity (IQG)}, \textsf{Friedrich-Alexander-Universit\"at Erlangen-N\"urnberg (FAU)},\protect\\ \textsf{91058 Erlangen, Germany}}
\affil[3]{\textsf{Department of Physics, South China University of Technology,}
\protect\\ \textsf{510641 Guangzhou, China}}

\date{{\small\sf \today}}                    
\usepackage{hyperref}

\begin{document}

\maketitle

{\sf

\begin{abstract}

Loop quantum gravity in its Hamiltonian form relies on a connection formulation of the gravitational phase space with three key properties: 1.) a compact gauge group, 2.) real variables, and 3.) canonical Poisson brackets. In conjunction, these properties allow to construct a well defined kinematical quantization of the holonomy flux-algebra on top of which the remaining constraints can be implemented. While this idea has traditionally been mainly used for Einstein gravity, any gravitational theory with the above properties can be accommodated. In this paper, we are going to review three strands of work building on this observation, namely the study of higher-dimensional loop quantum gravity, supersymmetric extensions of loop quantum gravity, as well as the quantization of modified gravitational theories.

\end{abstract}

}
\newpage

\section{Introductory remarks}

Hamiltonian loop quantum gravity relies at its core on a kinematical construction leading to the Ashtekar-Lewandowski Hilbert space \cite{AshtekarRepresentationsOfThe, AshtekarRepresentationTheoryOf, AshtekarDifferentialGeometryOn, AshtekarProjectiveTechniquesAnd, MarolfOnTheSupport, AshtekarQuantizationOfDiffeomorphism}. This Hilbert space carries a representation of the classical holonomy-flux algebra that provides a point-splitting subset of phase space functions, thus leading to a kinematical quantization. At this point, there are still constraints to be dealt with via Dirac quantization in order to compute the dynamics. A crucial input at the classical level is a formulation of general relativity (GR) in terms of connection variables with certain properties discussed below.
In the seminal papers, mainly the gravitational degrees of freedom were discussed, but matter field can be included in analogy to ideas from lattice gauge theory \cite{ThiemannKinematicalHilbertSpaces, ThiemannQSD5}.

While treating general relativity in 3+1 dimensions along with rather arbitrary matter fields is already a quite satisfactory situation, one may ask to which extend one can apply the techniques developed in loop quantum gravity to other gravitational theories. Motivation to do so can be given as follows:
\begin{enumerate}
	\item Higher dimensional gravitational theories are mainly interesting from the point of view of making contact with string theory, see e.g. \cite{PolchinskiBook1, PolchinskiBook2}, which requires 10, 11, or 26 dimensions respectively for consistent formulations, as well as applications of the gauge / gravity duality, see e.g. \cite{AmmonGaugeGravityDuality}, where one would e.g. like to consider loop quantum gravity in 4+1-dimensional anti-de Sitter spacetimes to make statements about 3+1-dimensional gauge theories, see e.g. \cite{BodendorferHolographicSignaturesOf, BodendorferHolographicSignaturesOf2}.
	\item Historically, supersymmetric gravitational theories have been considered due to their enhanced UV-behaviour as well as for unification purposes, see e.g. \cite{NieuwenhuizenSupergravity}. Nowadays, they are mainly relevant for string theory and holography, where they appear as low energy theories and the relevant gravitational theories respectively, see e.g. \cite{OrtinGravityAndStrings}. Next to contact with string theory and holography as above, it is interesting to consider supersymmetric gravitational theories by themselves and to investigate what can be learned about the quantum dynamics due to the more complicated algebra of constraints \cite{TeitelboimSupergravityAndSquare}.
	\item Modified gravity theories play an important role in cosmology \cite{So}. Observations strongly implied that our universe is currently undergoing a period of accelerated expansion and is usually referred to as the "dark energy" issue. The origin of current cosmic acceleration is one of the biggest challenges to modern physics. This issue is hard to accounted in General Relativity(GR) framework. Hence it is reasonable to consider the possibility that GR is not a valid gravity theory at the cosmological scale and should be modified. If nature is indeed described by such a modified gravitational theory, a loop quantum gravity type quantization of such theories needs to be investigated as well.
\end{enumerate}

As mentioned in the abstract, the three key properties that a connection formulation of a gravitational theory should enjoy in order to directly apply the quantization methods of loop quantum gravity are 1.) a compact gauge group, 2.) real variables, and 3.) canonical Poisson brackets. From the canonical brackets, one obtains the standard form of the holonomy-flux algebra, in particular commutativity of holonomies among themselves and the action of fluxes as ``grasping operators'' which split holonomies at the intersection points. On this algebra, the Ashtekar-Lewandowski functional is positive, linear and normalized, thus providing a Hilbert space representation. Compactness of the gauge group allows to implement the continuum limit by accounting for cylindrical consistency as well as a rigorous quantization of the constraints. Moreover, the Hilbert space representation yields a useful basis in terms of spin-network states. Finally, real variables ensure that the reality conditions are trivially implemented.

However, as we will see explicitly in Sec. \ref{sec:SUSYconstraint} below, in the context of supergravity, it turns out that using real variables leads to a rather complicated form of the constraint operators, in particular the so-called SUSY constraint operator, making direct physical applications and predictions almost impossible. This changes drastically if complex variables are used instead, which recover some of the underlying supersymmetry of the theory and thus simplify the constraints. These observations allow for direct physical applications such as in the context of cosmology and black holes. The prize to pay, however, is the non-compactness of the resulting gauge group and the question on how to implement reality conditions making quantization arguments more subtle.

\section{Higher dimensions}\label{sec:HigherDimensions}

In this part of the article, we review the papers \cite{BTTVIII, BTTI, BTTII, BTTIII, BTTIV, BTTV, BTTVI, BTTVII}, where the higher-dimensional connection formulation is developed and the existing quantization techniques have been extended where necessary. Some aspects of boundaries that have been discussed in \cite{BTTXII, BI, BII} are also mentioned. We omit more recent work on, e.g., coherent states \cite{LongPerelomovTypeCoherent}, the simplicity constraint \cite{LongCoherentIntertwinerSolution}, and polytopes \cite{LongPolytopesInAll} for brevity.

In short, it was possible to extend the quantisation methods of loop quantum gravity to higher dimensions. The key result was a connection formulation of higher-dimensional general relativity with the properties discussed in the introduction \cite{BTTI, BTTII}. Based on this, the quantization procedure could be straight-forwardly adapted, geometric operators for (spatial codimension one) area and volume could be defined,
and a quantization of the Hamiltonian constraint could be constructed \cite{BTTIII}. Moreover, extensions for most matter fields, including those of several interesting supergravity theories could be constructed \cite{BTTIV, BTTVI, BTTVII}. A key new ingredient, the simplicity constraint, was discussed in \cite{BTTV}.

\subsection{Connection formulation of higher dimensional GR}

\subsubsection{General considerations}

As mentioned in the introduction, we want to quantize a Hamiltonian formulation of GR in terms of a connection and its conjugate momentum. As a starting point for the derivation, we use the Arnowitt-Deser-Misner (ADM) formulation of GR \cite{ArnowittTheDynamicsOf} in terms of metric variables. As usually, we assume that our $(D+1)$-dimensional spacetime manifold $\mathcal M$ foliates as $\mathbb R \times \Sigma$, where $\Sigma$ is a $D$-dimensional Cauchy surface. On $\Sigma$, the spacetime metric $g_{\mu \nu}$, $\mu, \nu = 0,\ldots,D$ induces a Riemannian metric $q_{ab}$, $a,b = 1,\ldots,D$. The extrinsic curvature of $\Sigma$ in $\mathcal M$ is denoted by $K_{ab}$. We use the convention
\be
	S_{\text{EH}} = \frac{1}{2 \kappa} \int_{\mathcal M} \sqrt{-g} R(g) d^{D+1}x
\ee
for the Einstein-Hilbert action, which leads to the ADM Poisson bracket
\be
	\{ q_{ab}(x), P^{cd}(y) \} = \delta^{(D)}(x-y) \delta_{(a}^c \delta_{b)}^d
\ee
with $P^{ab} = \frac{1}{2 \kappa} \sqrt{q}(K^{ab}-q^{ab} K^{cd} q_{cd})$. We will set $\kappa = 8 \pi G = 1$ in the following. The Hamiltonian of the theory is given by a sum of the Hamiltonian constraint $S(x)$ and the spatial diffeomorphism constraint $V_a(x)$, smeared against the lapse function $N(x)$ and shift vector $N^a(x)$, giving
\be
	H_{\text{ADM}} = \int_\Sigma \left( N(x) S(x) + N^a(x) V_{a}(x)\right) d^D x \text{.}
\ee
For later reference, we explicitly state
\be \label{eq:SLE}
	S=-\frac{s}{\sqrt{\det(q)}}[q_{ac} q_{bd}-\frac{1}{D-1} q_{ab} q_{cd}]P^{ab} P^{cd}-\sqrt{\det(q)} R^{(D)} \text{,}
\ee
where $s$ represents the signature of spacetime as $-1$ for Lorentzian and $+1$ for Euclidean.

The technical means for obtaining a connection formulation starting from the ADM formulation is that of a phase space extension. The basic idea is to postulate a new, larger phase space, subject to additional first class constraints. This phase space should be coordinatised by a connection $A_{a \alpha}$, $\alpha$ being a Lie algebra index in a Lie group $\mathcal G$ and its canonically conjugate\footnote{We exclude the possibility of having non-canonical brackets between $A_{a \alpha}$ and $\pi^{a \alpha}$, since this would lead to a non-standard holonomy-flux algebra, on which is the Ashtekar-Lewandowski measure might not lead to a positive linear functional.} momentum $\pi^{a\alpha}$.  The list of constraints should include a Gau{\ss} law $G^{\alpha}(x) :=D_a \pi^{a \alpha}(x) = 0$, which generates local $\mathcal G$-gauge transformations. Furthermore, we need explicit expressions of the ADM variables $q_{ab}$ and $P^{ab}$ in terms of $A_{a\alpha}$ and $\pi^{a\alpha}$. Then, what one has to proof is that the ADM Poisson brackets are reproduced up to the new constraints, that is
\be
	\left\{ q_{ab}[A, \pi](x), P^{cd}[A, \pi](y) \right\} \approx \delta^{(D)}(x-y) \delta_{(a}^c \delta_{b)}^d \text{,} \label{eq:ADMOfAAndPi}
\ee
where $\approx$ denotes a weak equality, that is equality up to constraints. The remaining Poisson brackets have to vanish. Finally, one expresses $S$ and $V^a$ in terms of $A_{a\alpha}$ and $\pi^{a\alpha}$. The physics of the new formulation is now equivalent to the ADM formulation, and the extra degrees of freedom constitute a gauge redundancy in the description.

A priori, there is a vast landscape of possible connection formulations and one would like to select an appropriate one. It turns out when considering only the Gau{\ss} law as a new constraint and a vielbein-type construction of the connection variables, the possible gauge groups are restricted to SO$(3)$ and SO$(1,2)$, or their respective universal covers \cite{BTTI}. The corresponding canonical variables are then either the drei-bein and its conjugate connection in 2+1 gravity, or the Ashtekar-Barbero variables \cite{AshtekarNewVariablesFor, BarberoRealAshtekarVariables} in $3+1$ dimensions.

While the quantisation and solution of the Gau{\ss} law is straight forward in loop quantum gravity, we would furthermore like to demand that the quantisation of the additional constraints is ``well behaved''. For example, if the additional constraints could be expressed solely in terms of fluxes, that is $\pi^{a\alpha}$ smeared over a spacetime codimension 2 surface, they would leave the graph on which spin networks are defined invariant. This in turn means that the kernel of these constraints can be constructed at the level of individual spin networks, i.e. by restricting the allowed representations and invariant maps. In other words, it would be a purely group theoretical exercise, and not involve (infinite) superpositions of spin networks, which might be non-normalisable in the kinematical scalar product.


While we cannot exclude the existence of further connection formulations with the above mentioned properties, the only two known at present are motivated by the Palatini action. Here, the gauge group is either SO$(D+1)$ or SO$(1,D)$, starting with Euclidean or Lorentzian spacetime signature. As we will see however later, at the level of the Hamiltonian theory, both formulations can be used for either Euclidean or Lorentzian gravity, since the signature of spacetime is only encoded in a relative sign in the Hamiltonian constraint. This, at first unintuitive property, becomes clear when considering that in both formulations, the ADM variables are a Riemannian D-metric and its conjugate momentum. A phase space extension performed in either the Euclidean or Lorentzian thus has to be valid also in the other theory. For quantisation purposes, the compact group SO$(D+1)$ is strongly preferred by the currently available techniques in loop quantum gravity and we will choose this formulation.


\subsubsection{The Palatini action }

\label{sec:PalatiniAction}

We start our motivation for the choice of connection variables by performing a canonical $D+1$-decomposition of the Lorentzian Palatini action
\be
	S_{\text{Palatini}} = - \frac{1}{2} \int_\mathcal{M} e \, e^{\mu I} e^{\mu J} F_{\mu \nu IJ}(A) d^{D+1}x =: \int_\mathcal{M} \Sigma^{IJ}(e) \wedge F_{IJ}(A)  \text{,} \label{eq:PalatiniAction}
\ee
where $e^\mu_I$ is a $(D+1)$-bein, $e = \det e_{\mu I}$, $A_{aIJ}$ an SO$(1,D)$ connection, and $F_{\mu \nu IJ}$ its curvature. The global sign of the action is chosen to agree with the conventions in \cite{BTTII}, where the following calculations are detailed. Assuming $\mathcal M = \mathbb R \times \Sigma$, with $\Sigma$ compact without boundary, we have
\be
	  S_{\text{Palatini}} =\int dt \int_\Sigma d^{D}x \, \left( \frac{1}{2} \pi^{aIJ} \mathcal{L}_T A_{aIJ} - N \mathcal{S} - N^{a} {V}_a - \frac{1}{2} \lambda_{ IJ} G^{IJ} \right) \text{,}
\ee
where $\pi^{aIJ} = 2 n^{[I} \sqrt{q} e^{a|J]}$, with $n^I = n^\mu e_{\mu}^I$ being the internal version of the timelike unit normal $n^{\mu}$ to $\Sigma$, and $\mathcal{L}_T$ denoting the Lie derivative with respect to the time evolution vector field $T^{\mu} = N n^\mu + N^{\mu}$. $S$ and $V_a$ are again the Hamiltonian and spatial diffeomorphism constraints, and $G^{IJ} = D_a \pi^{aIJ} = \partial_a \pi^{aIJ} + [A, \pi^a]^{IJ}$ is the Gau{\ss} constraint with smearing functions $\lambda_{IJ}$. Starting from here, one can perform a canonical analysis following Dirac \cite{DiracLecturesOnQuantum}. While this can be done in all detail \cite{BTTII}, the problem is that one would have to introduce additional variables, the momenta to $A_{aIJ}$ and $e^{aI}$. However, we are interested in a formulation in terms of variables $A_{aIJ}$ and $\pi^{aIJ}$ only. This can be accomplished by the following observation: if we just use $A_{aIJ}$ and $\pi^{aIJ}$ as conjugate variables, forgetting about the fact that $\pi^{aIJ}$ decomposes as $2 n^{[I} \sqrt{q} e^{a|J]}$, we would make an error, since we would have too many degrees of freedom. However, we also notice that $S$, $V_a$, and $G^{IJ}$ can be expressed purely in terms of $A_{aIJ}$ and $\pi^{aIJ}$ \cite{BTTII}, keeping in mind that $\pi^{aIJ} = 2 n^{[I} \sqrt{q} e^{a|J]}$. This leads us to the conclusion that if we can write a constraint purely in terms of $A_{aIJ}$ and $\pi^{aIJ}$ such that $\pi^{aIJ} = 2 n^{[I} \sqrt{q} e^{a|J]}$ on the constraint surface, we can use the phase space coordinatised by $A_{aIJ}$ and a ``generic'' momentum $\pi^{aIJ}$, subject to this additional constraint.

Surprisingly, it is possible to write down such a constraint in the simple form
\be
	S^{ab\,IJKL} := \pi^{a[IJ} \pi^{b|KL]} = 0 \text{,} \label{eq:SimplicityConstraint}
\ee
which is closely related to the Plebanski formulation of GR \cite{PlebanskiOnTheSeparation}. In loop quantum gravity, this constraint is known as the (quadratic\footnote{One can also write down a linear version of this constraint, see for example \cite{EngleLQGVertexWith, BTTVI}. The quadratic constraint in $3+1$ dimensions suffers from the problem of admitting a topological sector, which has to be excluded by hand. See also \cite{EngleAProposedProper} for further discussion on related subtleties.}) simplicity constraint, and first appeared in the Barrett-Crane spinfoam model \cite{BarrettRelativisticSpinNetworks}. It satisfies the above requirement of being expressible in terms of fluxes only and indeed translates in to a restriction of the allowed group representation labels in the quantum theory, as we will discuss later.

With this constraint added to the list of constraints of the theory, we now performa canonical analysis. As detailed in \cite{BTTII}, this leads to a new constraint which originates from demanding the simplicity constraint to be preserved by the Hamiltonian evolution. We will denote this constraint by $D^{ab\,IJKL}$, surpressing indices in what follows for notational simplicity. Its precise form doesn't matter for this paper, and we just remark that it sets a certain part of the torsion of $A_{aIJ}$ to zero. $D$ turns out to be a second class partner to $S$, which is unwanted, because it was our aim to impose $S^{ab\,IJKL}$ as a strong operator equation in the quantum theory. As is well known, second class constraints require different quantisation techniques, and $D$ furthermore depends on $A_{aIJ}$, leading to a complicated operator.

A way out of this problem is the technique of gauge unfixing \cite{MitraGaugeInvariantReformulationAnomalous, AnishettyGaugeInvarianceIn, VytheeswaranGaugeUnfixingIn}, which transforms a second class constrained system in an equivalent first class constrained system. It can be seen as the Hamiltonian analogue of the St\"uckelberg trick and essentially amounts to the physical equivalence of a theory with gauge freedom before and after gauge fixing. Using this technique, we can view $D$ as a gauge fixing condition for the simplicity constraint and remove it from the list of constraints. We note that $S$ trivially Poisson commutes with itself. The only modification this amounts to is to change the Hamiltonian constraint by essentially adding $D^2$ in such a way that it Poisson commutes with the simplicity constraint. On the constraint surface of the simplicity and Gau{\ss} constraints, the new ``gauge unfixed'' Hamiltonian constraint then reduces to its ADM version.

We thus conclude that we found a Hamiltonian formulation of GR with the following properties:
\begin{itemize}
	\item The phase space is of Yang-Mills type, coordinatised by the SO$(1,D)$-connection $A_{aIJ}$ and its conjugate momentum $\pi^{aIJ}$.
	\item The Poisson brackets are canonical, i.e. $\{A_{aIJ}(x), \pi^{bKL}(y)\}=2 \delta_a^b  \delta_{[I}^K \delta_{J]}^L \delta^{(D)}(x-y)$.
	\item The canonical variables are real.
	\item The theory is subject to the first class constraints $\mathcal S$, $V_a$, $G^{IJ}$, and $S^{ab\,IJKL}$.
	\item The Hamiltonian is a sum of the (smeared) first class constraints.
\end{itemize}
Up to the non-compact gauge group, we have thus found a suitable Hamiltonian connection formulation of GR in any dimension $D+1 \geq 3$. In $2+1$ dimension, the simplicity constraints are trivially zero and the formulation thus simplifies to the usual canonical formulation of the $2+1$-dimensional Palatini action.

\subsubsection{Compact internal group}

The only shortcoming of the above canonical formulation of GR which prevents us from applying the quantisation techniques on which loop quantum gravity in $3+1$ dimensions is based is the non-compactness of the gauge group, which enters the construction of the Hilbert space  by providing a normalisable Haar measure. From an aesthetical point of view, one can argue that having SO$(1,D)$ as an internal gauge group would better reflect the physics of the theory, due to its origin in the Palatini action, whereas its compact analogue SO$(D+1)$ results form the Euclidean theory. On the other hand, the internal gauge group is completely redundant for the physics at the classical level, since by solving the simplicity constraint and Gau{\ss} constraint, the theory can be reduced to the ADM phase space, which doesn't care from which internal gauge group it has originally been obtained. Its dynamics is solely governed by the remaining constraints, the Hamiltonian and spatial diffeomorphism constraints.

This  now leads us to the following observation: consider the Euclidean Palatini action and perform a canonical analysis as in the previous section. We obtain the same theory, up to the fact that the internal gauge group now is SO$(D+1)$ and the Hamiltonian constraint encodes the dynamics of the Euclidean theory. Upon solution of the simplicity and Gau{\ss} constraints, we obtain the Euclidean ADM phase space, whose only difference to the Lorentzian ADM phase space is the Hamiltonian constraint, more precisely a relative sign $s$ between the two terms in \eqref{eq:SLE}. It now follows that we can write down a theory based on the SO$(D+1)$ connection variables coming from the Euclidean Palatini action by modifying the Hamiltonian constraint such that it reduces to the Lorentzian ADM constraint upon solving the Gau{\ss} and simplicity constraints. This is indeed possible, as detailed in \cite{BTTI}.

Let us summarise our findings for now: We have a Hamiltonian theory based on an SO$(D+1)$ connection and its canonical momentum $\pi^{aIJ}$, subject to the canonical brackets
\be
	\{A_{aIJ}(x), \pi^{bKL}(y)\}=2 \delta_a^b  \delta_{[I}^K \delta_{J]}^L \delta^{(D)}(x-y) \text . \label{eq:BracketAPi}
\ee
The spatial metric $q_{ab}$ is encoded as
\be
	2 q q^{ab} = \pi^{aIJ} \pi^{b}_{IJ},
\ee
while the relation to the extrinsic curvature, and thus $P^{ab}$, is a little more involved. For it, we note that there exists a unique SO$(D+1)$ ``hybrid spin connection'' $\Gamma^{\text H}_{aIJ}$ with the property $D^{\Gamma^{\text H}}_a n^I = D^{\Gamma^{\text H}}_a e^b_I = 0$ \cite{PeldanActionsForGravity}, where $D^{\Gamma^{\text H}}_a$ acts on tensor indices with the Christoffel symbols. It can be expressed as a function of $\pi^{aIJ}$ only \cite{BTTI}. Now, $A_{aIJ}$ decomposes as
\be
	A_{aIJ} = \Gamma^{\text H}_{aIJ} (\pi) + K_{ab} \pi^{b}_{IJ} / \sqrt{q} + \bar{K}_{aIJ} \text{,}
\ee
where $\bar{K}_{aIJ}$, subject to $\bar{K}_{aIJ} n^I=0$, is pure gauge \cite{BTTI}. It now can be shown that \eqref{eq:ADMOfAAndPi} is satisfied up to the Gau{\ss} and simplicity constraints
\be
	G^{IJ}[\lambda_{IJ}] = \int_\Sigma \lambda_{IJ} D_a \pi^{aIJ} d^Dx, ~~~~~ S^{ab\,IJKL}[c_{ab IJKL}] := \int_\Sigma c_{ab IJKL} \pi^{a[IJ} \pi^{b|KL]} d^Dx= 0 \text{.} \label{eq:ConstraintsGaussSimplicity}
\ee
$c_{ab IJKL}$ has density weight $-1$. The spatial diffeomorphism constraint reads (up to a boundary term obtained from the York-Gibbons-Hawking \cite{} boundary term in the action)
\be
	V_a[N^a] =  \frac{1}{2} \int_\Sigma \pi^{aIJ} \mathcal L_{N} A_{aIJ} d^Dx \text , \label{eq:ConstraintDiff}
\ee
with $\mathcal L_N$ being the Lie derivative with respect to $N^a$. The form of the Lorentzian Hamiltonian constraint in these variables has been given in \cite{BTTI}.

Since we already allowed us to use a ``non-standard'' formulation to describe GR, we can check whether there is an additional possibility to modify the variables further. Indeed, a simple modification results form rescaling  the $K_{ab}$ part in $A_{aIJ}$ and (inversely so) the $\pi^{aIJ}$ by a non-zero real parameter $\beta$\footnote{The free parameter $\beta$ is analogous to the Barbero-Immirzi parameter $\gamma$ \cite{BarberoRealAshtekarVariables, ImmirziQuantumGravityAnd} known from 3+1 dimensions, however different in the sense that in $3+1$ dimensions, a two parameter family of connection formulations in terms of $\beta$ and $\gamma$ exists \cite{BTTII}, and $\gamma$ is restricted to $3+1$ dimensions only. Moreover, in even spacetime dimensions one can consider yet another modification of the connection variables related ot the $\theta$-ambiguity in QCD, see \cite{}.}. Then, the new relation to the ADM phase space is
\be
	\frac{2}{\beta^2} q q^{ab} = {}^{(\beta)}\pi^{aIJ} {}^{(\beta)}\pi^{b}_{IJ},~~~~	{}^{(\beta)}A_{aIJ} = \Gamma^{\text H}_{aIJ} ({}^{(\beta)}\pi) + \beta K_{ab} (\beta \, {}^{(\beta)}\pi^{b}_{IJ}) / \sqrt{q} + \bar{K}_{aIJ} \text{,}
\ee
with ${}^{(\beta)} \pi^{aIJ} = \pi^{aIJ}/\beta$, while \eqref{eq:BracketAPi} remains unchanged and in \eqref{eq:ConstraintsGaussSimplicity} and \eqref{eq:ConstraintDiff} the rescaled variables simply replace the old ones.

The free parameter $\beta$ plays an important role in the quantum theory. While classically it amounts to performing a canonical transformation and thus doesn't change the physics, it turns out that these canonical transformations cannot be implemented as unitary transformations in the quantum theory or even algebra automorphisms on the holonomy-flux algebra. We restricted to real $\beta$ in order to obtain a real connection formulation, whose reality conditions are implemented by the kinematical scalar product. However classically, complex values of $\beta$ would work equally well.



\subsubsection{Boundaries}

Boundaries have been widely investigated within loop quantum gravity, in particular for the purpose of computing black hole entropy, see the respective chapters in this book. In this article, we include a brief discussion about higher dimensions to show that the results from 3+1 dimensions directly generalize, see \cite{BTTXII, BI, BII} for details.

For the treatment of general boundaries, we will choose a complementary approach to the covariant canonical description and build on existing results in ADM-variables \cite{HawkingTheGravitationalHamiltonian}. As boundary condition, we choose to fix the induced metric on the boundary, which leads to the well-known York-Gibbons-Hawking boundary term. This boundary term then has to be taken into account in the $(D+1)$-split of the action, as is done in \cite{HawkingTheGravitationalHamiltonian}. It results into boundary terms in the spatial diffeomorphism and Hamiltonian constraints. Most relevant for us, the spatial diffeomorphism constraint, including its boundary contribution\footnote{We include the boundary term of the constraint in order for it to generate spatial diffeomorphisms also at the boundary. The surface area $A_H$ of a boundary slice is invariant under such spatial diffeomorphisms and it is thus physically sensible to mod them out also in the quantum theory for the purpose of computing black hole entropy, which is classically determined by $A_H$.}, is given by
\be
	V_a [N^a] = \int_\Sigma P^{ab} \mathcal L_N q_{ab} \text{,} \label{eq:ADMGeneratorSpatialDiff}
\ee
where the shift vector is restricted to satisfy $s_a N^a = 0$ on $H$ in order to preserve the boundary. The symplectic structure has no boundary contribution at the level of the ADM variables.

For the phase space extension to connection variables, we need to check whether it leads to a boundary contribution in the symplectic structure and in the constraints when expressed in a suitable form. At the level of the symplectic potential, it can be shown that
\be
	\int_\Sigma P^{ab} \delta q_{ab} \approx \frac{1}{2} \int_\Sigma {}^{(\beta)}\pi^{aIJ} \delta {}^{(\beta)}A_{aIJ} + \frac{1}{\beta} \int_{\partial \Sigma} n^I \delta {\tilde{s}}_I  + \delta (\ldots)\text{,} \label{eq:PhaseSpaceExtensionBoundary}
\ee
where $\approx$ means equality up to the simplicity constraint, leading to the conclusion that the usage of connection variables always leads to a boundary contribution to the symplectic structure. In fact, the boundary term results from the identity\footnote{The signs in \eqref{eq:PhaseSpaceExtensionBoundary} and \eqref{eq:ExtensionIdentity} are compatible since $s_a$ is chosen inward pointing in $\Sigma$.}
\be
	\frac{1}{2} {}^{(\beta)}\pi^{aIJ} \delta \Gamma^{\text H}_{aIJ}  - \frac{1}{\beta} \partial_a \left( E^{aI} \delta n_I \right) \approx \delta (\ldots) \text , \label{eq:ExtensionIdentity}
\ee
which needs to be invoked in the passage to our connection variables.
The total variation in \eqref{eq:PhaseSpaceExtensionBoundary} vanishes when computing the symplectic structure and thus the Poisson brackets.

The spatial diffeomorphism constraint can be rewritten in terms of the connection variables as
\be
	V_a [N^a] = \int_\Sigma P^{ab} \mathcal L_N q_{ab} \approx \frac{1}{2} \int_\Sigma  {}^{(\beta)} \pi^{aIJ} \mathcal L_N  {}^{(\beta)} A_{aIJ} + \frac{1}{\beta} \int_{\partial \Sigma} n^I \mathcal L_N \tilde s_I\text{.}
\ee
The weak equality here is up to a term proportional to the ``boost part'' $n_I G^{IJ}$ of the Gau{\ss} constraint. It thus generates Lie derivatives on both the bulk and boundary variables. In order to compute the gauge transformations of the Gau{\ss} constraint, we have to partially integrate to free the canonical variables from derivatives, resulting in
\be
	G^{IJ}[\lambda_{IJ}] = \int_{\Sigma} \lambda_{IJ} D_a  {}^{(\beta)}\pi^{aIJ} = -\int_{\Sigma}  {}^{(\beta)}\pi^{aIJ} D_a \lambda_{IJ} + \frac{2}{\beta} \int_{\partial \Sigma} \lambda_{IJ} n^I \tilde s^J \text{.}   \label{eq:GaussLaw}
\ee
Also here, we see that the local SO$(D+1)$ gauge transformations act on both the bulk and boundary variables.

\subsubsection{Holonomy-flux algebra}

The quantum theory which we will derive from the above variables has strong similarities to lattice QCD, with the main difference that the lattice will become a dynamical object in the quantum theory, i.e. one works, in a certain precise sense, on all lattices simultaneously. We start by recalling that for a canonical quantisation, we need to find a representation of a point-separating (unital) Poisson *-subalgebra of phase space functions on a Hilbert space. It is therefore our first task to find a subalgebra of phase space functions which admits such a representation. In making this choice, we are guided by the constraints of the theory, that is we would like to have an easy transformation property of the algebra elements at least under SO$(D+1)$ gauge transformations and spatial diffeomorphisms. Moreover, we would like to solve these constraints later on. The technical details of the following construction can be found in \cite{ThiemannModernCanonicalQuantum}, see also \cite{BTTIII} for a short summary focussing on higher dimensions.

From the SO$(D+1)$ gauge transformations, it suggests itself to use holonomies of $A_{aIJ}$, since these can be used to form Wilson loops, or more generally, spin networks. We denote them by
\be
	h^{\lambda}_e(A) = \mathcal P \exp \left( \int_e A_{aIJ}\, \tau_{\lambda}^{IJ} dx^a\right) \text{,}  \label{eq:HolonomyDefinition}
\ee
where $e$ is a path (edge of a spin network) in $\Sigma$, $\mathcal P$ the path ordering symbol, and $\tau_{\lambda}^{IJ}$ are generators of so$(D+1)$ in a representation labelled by $\lambda \in \mathbb N_0$\footnote{We will see later on that the representations in the quantum theory are labelled by a single non-negative integer $\lambda$. We thus refrain from introducing proper notation such $h^{\vec \Lambda}_e(A)$ for a holonomy in a general representation with heightest weight vector $\vec \Lambda$.}. All gauge invariant information about the connection is encoded in the set of all holonomies. Next, we define the fluxes
\be
	\pi^n_S = \int_S n_{IJ} \pi^{aIJ} \epsilon_{ab_1 \ldots b_{D-1}} dx^{b_1} \wedge \ldots \wedge dx^{D-1} \label{eq:Flux}
\ee
generalising the electric fluxes of Maxwell theory to a non-Abelian gauge group. $n_{IJ}$ is a smearing function and $S$ a $(D-1)$-dimensional surface.

Using \eqref{eq:BracketAPi}, we can now compute the Poisson bracket between holonomies and fluxes. Exemplarily, in the case that $e$ intersects $S$ transversally in the point $p \in \Sigma$, we find
\be
	\left\{h_e(A), \pi^n_S \right\} \propto \pm h_{e_1}(A) \tau^{IJ} n_{IJ}(p) h_{e_2}(A) \text{,} \label{eq:HolonomyFluxCommutator}
\ee
where $e = e_2 \circ e_1$ has been split at $p$. The sign in \eqref{eq:HolonomyFluxCommutator} depends on the intersection properties, that is the orientation of $e$ and $S$. Non-transversal and multiple intersection are detailed in \cite{ThiemannModernCanonicalQuantum} and we will neglect to discuss them here, as they are not required for the basic understanding of the quantisation procedure. The Poisson bracket between two holonomies vanishes trivially, while the Poisson bracket between two fluxes cannot, in general, be expressed in terms of holonomies and fluxes again\footnote{This results from the singular smearing in terms of holonomies and fluxes, see \cite{ThiemannModernCanonicalQuantum}.}. This is however not an obstacle for the following quantization.

On $\partial \Sigma$, we consider the phase space functions
\be
	L^{IJ} = \hat s_a \pi^{aIJ} = 2/\beta \, n^{[I} \tilde s^{J]} \text{,}	\label{eq:DefL}
\ee
smeared over surfaces $S$ in the same manner as \eqref{eq:Flux}. Using \eqref{eq:PhaseSpaceExtensionBoundary}, their Poisson bracket reads
\be
	\{ L^{IJ}_S, L^{KL}_{S'} \} = 4 \delta^{J][K} L^{L][I}_{S \cap S'} \text{,} ~~ \text{or} ~~ \{ L^{n}_S, L^{n'}_{S'} \} = -2 L^{[n,n']}_{S \cap S'}\label{eq:BracketLL}
\ee
in terms of smearing functions $n^{IJ}$ and $n'^{IJ}$.
The boundary variables are thus just the restrictions of the fluxes to $\partial \Sigma$.

We can now take the free algebra generated by all holonomies and fluxes and divide it by the commutator relations implied by \eqref{eq:HolonomyFluxCommutator} and its generalisations, that is $[A,B] = \hbar / i \{A,B\}$. This gives us a unital *-algebra $\mathfrak A$, which we now have to represent on a Hilbert space.

\subsection{Outline of quantization}

Most parts of the quantization procedure are analogous to the standard case in 3+1 dimensions up to replacing the gauge groups and we will not detail them here. Rather, we focus on the new aspects introduced by the simplicity constraint and its interplay with the area operator.


\subsubsection{The simplicity constraint}

We now proceed to implement the simplicity constraints \eqref{eq:SimplicityConstraint} as operator equations on the spin network basis states. Since only the smeared fluxes $\pi^n_{S}$, and not $\pi^{aIJ}(x)$, are well defined operators, we have to regularise \eqref{eq:SimplicityConstraint} in terms of fluxes. This has been done in detail in \cite{BTTIII}, and we recall the main results. The idea is to implement all of the constraints \eqref{eq:SimplicityConstraint} at a given point $x \in \Sigma$ by smearing each $\pi^{aIJ}$ over an arbitrary surface $S$ containing $x$, and then to compute the action of the operator
\be
	\hat \pi_S^{[IJ} \hat \pi_{S'}^{kl]}
\ee
on spin networks in the limit that the surfaces $S$ and $S'$ are shrunk to the point $x$. There are two non-trivial cases: $x$ is an inner point of an edge $e$, or $x$ is a vertex of the graph $\gamma$ on which the spin network is defined. The implementation on a vertex discussed here has been put forward in \cite{BTTV}, while the implementation on an edge follows directly from \cite{FreidelBFDescriptionOf}.

In the first case, a generic\footnote{$e$ is not contained in $S$ or $S'$ in a neighbourhood of $x$, otherwise the action vanishes.} choice of surfaces yields the condition
\be
	\tau^{[IJ}_{\vec \Lambda} \tau^{KL]}_{\vec \Lambda} =0 \label{eq:SimplicityTau}
\ee
with $\tau^{IJ}_{\vec \Lambda}$ being the generators of so$(D+1)$ in an irreducible representation labelled by the highest weight vector ${\vec \Lambda}$. This result follows directly from \eqref{eq:HolonomyFluxCommutator}, or visualising the $\pi^{aIJ}$ as derivative operators acting on the holonomies \eqref{eq:HolonomyDefinition}. It has been shown in \cite{FreidelBFDescriptionOf} that the condition \eqref{eq:SimplicityTau} on the generators of the Lie algebra so$(D+1)$ is equivalent to the restriction to a subset of representations with highest weight vector $\vec \Lambda = (\lambda, 0, \ldots, 0)$, $\lambda \in \mathbb N_0$. In the mathematical literature, these representations are called spherical, most degenerate, (completely) symmetric, or representations of class 1 (with respect to an SO$(D)$ subgroup). Within loop quantum gravity, they are usually denoted as ``simple'' representation, which originates from their relation to simple bi-vectors through the simplicity constraint. This result now justifies the notation $\tau^{IJ}_{\lambda}$ used in this paper, that is to denote the representations by $\lambda$ instead of $\vec \Lambda$.

The case of a vertex is more complicated, since the quantum simplicity constraints turn out to be non-commuting. Different strategies to deal with this problem in $3+1$ dimensions have been put forward in the literature, see \cite{BTTV} for an overview. We will consider here the dimension-independent solution proposed in \cite{BTTV}, which amounts to choosing a maximally commuting subset of vertex simplicity constraints, much in analogy with the choice of a maximal commuting subset of operators in quantum mechanics, or the strategy of gauge unfixing mentioned in section \ref{sec:PalatiniAction}. Technically, one first finds a set of ``basic'' simplicity constraints from considering all possible smearing surfaces, which turns out to be the most general set possible, that is the two fluxes each acting only on a single edge incident at $v$ \cite{BTTIII}. Next, one proposes a commuting subset of vertex simplicity constraints and shows that adding a new vertex simplicity constraint spoils commutativity \cite{BTTV}. The result of this procedure is rather intuitive: in a given recouping scheme, which is in one-to-one correspondence with the choice of maximally commuting subset, the intertwining representations are all simple. We call such an intertwiner simple (with respect to a given recoupling scheme). We note that such an intertwiner is in general not simple with respect to another recoupling scheme, as follows from the general decomposition
\cite{GirardiKroneckerProductsFor, GirardiGeneralizedYoungTableaux}
\be
	(\lambda_1,0,...,0) \otimes (\lambda_2,0,...,0) = \sum_{k = 0}^{\lambda_2}\sum_{l=0}^{\lambda_2-k} (\lambda_1 + \lambda_2 - 2k - l,l,0,...,0) \hspace{5mm} (\lambda_2 \leq \lambda_1) \label{eq:DecompositionSimpleSimple}
\ee
However, the dimension of the intertwiner space is independent of the choice of recoupling scheme, with respect to which the intertwiner is simple.
We remark that there exists an intertwiner which is simple with respect to every recouping scheme, known as the Barrett-Crane intertwiner in $3+1$ dimensions \cite{BarrettRelativisticSpinNetworks}. Its SO$(D+1)$ analogue has been constructed in \cite{FreidelBFDescriptionOf}.

It has not been checked explicitly if the operation of the quantum Hamiltonian constraint preserves any chosen simple intertwiner, or whether it needs to be modified to do so, e.g. by a gauge unfixing procedure.


\subsubsection{The area operator}\label{sec:area higher D}

We are now in the position to address the question of what happens to the geometrical properties of certain classical objects in the quantum theory. The so far best understood geometric operator in loop quantum gravity is the area operator, where area always refers to surface a spatial codimension 1. The area operator was originally proposed in \cite{SmolinRecentDevelopmentsIn}, see also \cite{BTTIII} for an account tailored to our situation. We will shortly sketch its derivation and compute its action on holonomies.

Classically, the area of a surface $S$ is given by
\be
	A[S] = \int_S d^{D-1}x \, \sqrt{\det h_{\alpha \beta}} \text{,}
\ee
where $h_{\alpha \beta}$, $\alpha, \beta = 1, \ldots,D-1$ is the induced metric on $S$, as inherited form $q_{ab}$. We can rewrite this expression as
\be
	A[S] = \int_S d^{D-1}x \, \sqrt{\hat n_a \hat n_b \, q q^{ab}} = \int_S d^{D-1}x \, \sqrt{\frac{\beta^2}{2} \hat n_a \hat n_b \, {}^{(\beta)} \pi^{aIJ}  {}^{(\beta)}\pi^{b} {}_{IJ} } \text{,}
\ee
where $\hat n_a = \epsilon_{a \beta_1 \ldots \beta_{D-1}} \epsilon^{\beta_1 \ldots \beta_{D-1}}/(D-1)!$ is the properly densitized conormal on $S$. As a next step, we need to express the area in terms of fluxes. To do this, we note that the above integral is defined as the limit of Riemann sum approximations. Using the decomposition $S = \cup_i S_i$, a set of points $x_i \in S_i$, and denoting by $|S_i|$ the coordinate volume of $S_i$, it follows that
 \be
	A[S] = \lim_{S_i \rightarrow 0} \sum_i |S_i| \sqrt{\frac{\beta^2}{2} \hat n_a(x_i) \hat n_b (x_i) {}^{(\beta)}\pi^{aIJ}(x_i) {}^{(\beta)}\pi^{b} {}_{IJ}(x_i) } = \lim_{S_i \rightarrow 0} \sum_i  \sqrt{\frac{\beta^2}{2}  \pi^{aIJ}_{S_i} \pi^{b} {}_{S_i IJ} } \text{.} \label{eq:AreaOperatorClassical}
\ee
We can thus quantise the area operator by substituting flux operators in \eqref{eq:AreaOperatorClassical} and appealing to the spectral theorem to define the square root\footnote{The operator under the square root is essentially self-adjoint and non-negative \cite{ThiemannModernCanonicalQuantum, BTTIII}.}. We study the action of this operator on a holonomy of an edge $e$ intersecting $S$ transversally, as this is enough for e.g. entropy computations. Splitting $e = e_2 \circ e_1$ at the intersection point, it follows that \cite{BTTIII}
\be
	\hat A[S] \, h^\lambda_{e}(A) =  8 \pi G \hbar \, \sqrt{\beta^2} \, h^\lambda_{e_2}(A) \sqrt{-2 \tau_\lambda^{IJ} \tau_{\lambda IJ}} \,h^\lambda_{e_1}(A) = 8 \pi G \hbar \, \sqrt{\beta^2} \, \sqrt{\lambda(\lambda+D-1)} \,h^\lambda_{e}(A) \text{,}
\ee	
that is the action of the operator is diagonal on holonomies, and proportional to the quadratic Casimir of SO$(D+1)$.

\subsection{Brief comparison in the case $D=3$}

For $D=3$, it is instructive to compare our results with the standard quantization based on Ashtekar-Barbero variables. The geometric structure of holonomies and fluxes is the same in both cases. The main difference are the gauge groups SU$(2)$ and SO$(4)$ respectively, as well as the additional simplicity constraint in the latter case. From the above discussion, we see that holonomies obeying the quantum simplicity constraint are labelled by a non-negative integer $\lambda$, which maps to $2 j$ when compared to the SU$(2)$ representation label $j$ \cite{BTTIII, BI}. In this case, the area operator eigenvalues agree up to a global constant that can be absorbed into $\beta$, the analogue of the Barbero-Immirzi paramter in the dimension-independent case. With the same mapping applied to the recoupling spins, also the intertwiners in both cases can be mapped onto each other, showing that the kinematical structure is very similar \cite{BTTV}.

On the other hand, a dynamical comparison has not been performed so far. For this, the action of qualitatively new terms in the Hamiltonian constraint that arise in order to retain a classical first class algebra \cite{BTTI, BTTII} needs to be studied in detail.

\section{Supersymmetry}
In this section, we want to review the results of the papers \cite{Eder:2022gge,Eder:2022mft,Eder:2021rgt,Eder:2020uff,Eder:2021ans} dealing with the question on how to combine LQG with the concept of supersymmetry (SUSY) playing a major role in supergravity (SUGRA) and superstring theory. Supersymmetry is a new kind of symmetry that arose in the context of the famous results of Coleman-Mandula \cite{Coleman:1967ad} and Haag-Lopuszanski-Sohnius \cite{Haag:1974qh} who were looking for symmetries of interacting
QFTs that can have a nontrivial mixture with spacetime symmetries. As a consequence, it follows that supersymmetry interrelates both bosonic and fermionic degrees of freedom, that is, both force and matter particles, and therefore seems to be a natural candidate for the search for a unified field theory. Trying to combine LQG with SUSY thus also brings LQG closer to ideas of unification. Another important motivation for following such a program is the desire to be able to build a bridge and find relations between the various approaches to quantum gravity. The first step to achieve this goal is to apply LQG techniques to supergravity. In \cite{BTTVI,BTTVII,BTTVIII}, the quantization of the (minimal) supersymmetric 11-dimensional SUGRA in the framework of LQG is performed using the connection variables in higher dimensions as discussed in the previous section. Here, we want to focus on (extended) SUGRA in four spacetime dimensions. To this end, we first review the Holst-Macdowell-Mansouri action of $\mathcal{N}=1,2$ AdS SUGRA in $D=4$ and subsequently outline the quantization of the so-called SUSY constraint using standard LQG techniques. After that, an alternative approach will be discussed using chiral variables unraveling an enlarged gauge symmetry which allows to keep SUSY more manifest in the classical as well as in quantum theory. Based on this observation, a quantization of the theory will be proposed. Finally, applications in the context of quantum supersymmetric black holes will be discussed.

\subsection{The Holst-MacDowell-Mansouri action for AdS-supergravity with boundaries}\label{sec:HolstMM}
The Holst-like extension of $\mathcal{N}$-extended pure Poincar\'e SUGRA actions in $D=4$ have been discussed in \cite{Kaul:2007gz}. Here, following \cite{Eder:2021rgt,Eder:2022mft}, we want to derive it in the case of a nontrivial cosmological constant $\Lambda=-3/L^2$ with $L$ the so-called anti-de Sitter radius using a geometric approach to SUGRA also known as the Castellani-D'Auria-Fr\'e approach \cite{DAuria:1982uck,Castellani:1991et,Castellani:2018zey,Andrianopoli:2014aqa,Andrianopoli:2020zbl}. In mathematical terms, this is based on the concept of a so-called super Cartan geometry (see \cite{Eder:2022mft,Eder:2021rgt,Eder:2021ans}). This also allows a discussion about the boundary theory and, in particular, its compatibility with local supersymmetry.

Pure AdS (Holst-)supergravity with $\mathcal{N}$-extended supersymmetry ($\mathcal{N}=1,2$) can be described in a geometric way in terms of a gauge field $\mathcal{A}$ also referred to as the super Cartan connection\footnote{In mathematical terms, this means that AdS (Holst-)supergravity for $\mathcal{N}=1,2$ is described in terms of a super Cartan geometry modeled on the super Klein geometry $(\mathrm{OSp}(\mathcal{N}|4),\mathrm{Spin}^+(1,3)\times\mathrm{SO}(\mathcal{N}))$ (see \cite{Eder:2022mft,Eder:2021rgt,Eder:2021ans}) for more details.} which takes values in the supersymmetric generalization of the anti-de Sitter algebra given by the orthosympletic superalgebra $\mathfrak{osp}(\mathcal{N}|4)$. This superalgebra is generated by bosonic generators $(P_I,M_{IJ},T^{rs})$ corresponding to infinitesimal spacetime translations, Lorentz-transformations and $R$-symmetry transformations, respectively, as well as $4\mathcal{N}$ fermionic generators ($Q^r_{\alpha}$) which combine to $\mathcal{N}$ Majorana spinors. These satisfy the following graded commutation relations
\begin{align}
[M_{IJ},Q^r_{\alpha}]&=\frac{1}{2}Q_{\beta}^r\tensor{(\gamma_{IJ})}{^{\beta}_{\alpha}}\label{A.eq:I.2.3.19}\\
[P_I,Q^r_{\alpha}]&=-\frac{1}{2L}Q^r_{\beta}\tensor{(\gamma_I)}{^{\beta}_{\alpha}}\label{A.eq:I.2.3.20}\\
[T^{pq},Q_{\alpha}^r]&=\frac{1}{2L}(\delta^{qr}Q_{\alpha}^p-\delta^{pr}Q_{\alpha}^q)\label{A.eq:I.2.3.21}\\
[Q_{\alpha}^r,Q_{\beta}^s]=\delta^{rs}\frac{1}{2}(C\gamma^I)_{\alpha\beta}P_I+&\delta^{rs}\frac{1}{4L}(C\gamma^{IJ})_{\alpha\beta}M_{IJ}-C_{\alpha\beta}T^{rs}
\label{A.eq:I.2.3.22}
\end{align}
The super Cartan connection can be decomposed in the following way
\begin{equation}
\mathcal{A}=e^IP_I+\frac{1}{2}\omega^{IJ}M_{IJ}+\frac{1}{2}\hat{A}_{rs}T^{rs}+\Psi_r^{\alpha}Q_{\alpha}^r
\label{D.eq:5.1.0}
\end{equation}
with $e^I$, $\omega^{IJ}$ and $\hat{A}_{rs}$ the co-frame, spin-connection and $\mathfrak{so}(\mathcal{N})$-gauge field, respectively, as well as $\Psi_r^{\alpha}$ the spin-3/2 Rarita-Schwinger fields.
This connection can be used in order to formulate a Yang-Mills-type action principle for Holst-supergravity. To this end, one introduces a $\beta$-deformed inner product $\braket{\cdot\wedge\cdot}_{\beta}$ on $\mathfrak{g}\equiv\mathfrak{osp}(\mathcal{N}|4)$-valued 2-forms on the underlying spacetime manifold $M$ with $\beta\in\mathbb{R}\setminus\{0\}$ the Barbero-Immirzi parameter via
\begin{align}
    \braket{\cdot\wedge\cdot}_{\beta}:\,\Omega^2(M,\mathfrak{g})\times\Omega^2(M,\mathfrak{g})\rightarrow\Omega^4(M),\quad(\omega,\eta)\mapsto\mathrm{str}(\omega\wedge\mathbf{P}_{\beta}\eta)
		\label{eq:deformedIP}
\end{align}
with ``$\mathrm{str}$'' denoting the $\mathrm{Ad}$-invariant supertrace on $\mathfrak{g}$ and $\mathbf{P}_{\beta}$ a $\beta$-dependent operator on $\Omega^2(M,\mathfrak{g})$ which for instance in the $\mathcal{N}=1$ case is given by
\begin{equation}
\mathbf{P}_{\beta}:=\underline{\boldsymbol{0}}\oplus\mathcal{P}_{\beta}\oplus\mathcal{P}_{\beta}\quad\text{ with }\quad\mathcal{P}_{\beta}:=\frac{\mathds{1}+ i\beta\gamma_5}{2\beta}
\label{D.eq:3.3}
\end{equation}
Using this inner product, the so-called Holst-MacDowell-Mansouri action of $\mathcal{N}$-extended pure AdS Holst-supergravity takes the form
\begin{equation}
S^{\beta,\mathcal{N}}_{\text{H-MM}}(\mathcal{A})=\frac{L^2}{\kappa}\int_{M}{\braket{F(\mathcal{A})\wedge F(\mathcal{A})}_{\beta}}
\label{D.eq:6.4}
\end{equation}
with $F(\mathcal{A})=\mathrm{d}\mathcal{A}+\frac{1}{2}[\mathcal{A}\wedge\mathcal{A}]$ the curvature of the super Cartan connection $\mathcal{A}$. Expanding \eqref{D.eq:6.4} in the components of the curvature, it follows that this action indeed yields the Holst-action of $\mathcal{N}$-extended AdS supergravity in the bulk together with certain boundary terms. As shown in \cite{Eder:2021rgt} based on arguments developed in \cite{Andrianopoli:2014aqa,Andrianopoli:2020zbl} using standard variables, the boundary terms arising from \eqref{D.eq:6.4} are indeed unique if one imposes local SUSY invariance of the full action at the boundary.

More precisely, in the Castellani-D'Auria-Fr\'e approach, local SUSY transformations are regarded as infinitesimal superdiffeomorphisms $\epsilon$ along the odd directions of an underlying supermanifold satisfying $i_{\epsilon}e^I=0$. One can then show that the Lagrangian $\mathscr{L}_{\text{full}}$ of the action \eqref{D.eq:6.4} is invariant under local SUSY transformations provided that, at the boundary $\partial M$, the condition \cite{Andrianopoli:2014aqa}
\begin{equation}
    (\iota_{\epsilon}\mathscr{L}_{\mathrm{full}})|_{\partial M}=0
    \label{eq.BoundaryCondition}
\end{equation}
is satisfied. It can then be shown that the boundary terms as arising from \eqref{D.eq:6.4} are indeed the uniquely fixed by this condition.

\subsection{Quantization of the SUSY constraint}\label{sec:SUSYconstraint}
The canonical analysis of the Holst action of $\mathcal{N}=1$ SUGRA in $D=4$ in the case of a vanishing cosmological constant has been first discussed in \cite{Sawaguchi:2001wi,Tsuda:1999bg}. In the case of higher dimensions using the Ashtekar-Barbero-type variables as introduced in section \ref{sec:HigherDimensions}, this has been considered in \cite{BTTVI,BTTVII,BTTVIII}. Here, we follow \cite{Eder:2020uff} where a consistent treatment of the reality conditions of the fermionic variables has been included. Expanding \eqref{D.eq:6.4} in the case $\mathcal{N}=1$ it follows that the action modulo boundary terms takes the form
\begin{align}
S_{\text{H-MM}}^{\beta,\mathcal{N}=1}=\frac{1}{2\kappa}\int_{M}\mathrm{d}^4x\,&\left(e e_I^{\mu}e_J^{\nu}\left[\tensor*{F(\omega)}{*_{\mu\nu}^{IJ}}-\frac{1}{2\beta}\tensor{\epsilon}{^{IJ}_{KL}}\tensor*{F(\omega)}{*_{\mu\nu}^{KL}}\right]+2\tensor{\epsilon}{^{\mu\nu\rho\sigma}}\bar{\psi}_{\mu}\gamma_{\sigma}\mathcal{P}_{\beta}D^{(\omega)}_{\nu}\psi_{\rho}\right.\label{C.eq:2.11}\\
&\left.-e\frac{1}{L}\bar{\psi}_{\mu}\gamma^{\mu\nu}\psi_{\nu}+\frac{6}{L^2}e\right)\nonumber
\end{align}
where $e:=\det(e^I_{\mu})$. By performing a 3+1 decomposition, i.e., by assuming that the spacetime $M$ is globally hyperbolic and thus accordingly splits in the form $M=\mathbb{R}\times\Sigma$ with $\Sigma$ a $3D$ spacelike Cauchy hypersurface, one finds that \eqref{C.eq:2.11} can be written in the following way
\begin{align}
S_{\text{H-MM}}^{\beta,\mathcal{N}=1}=\int_{\mathbb{R}}\mathrm{d}t\int_{\Sigma}\mathrm{d}^3x\,&\left(\frac{1}{\kappa\beta}E_i^{a}\mathcal{L}_T\tensor*[^{\beta\!\!}]{A}{_a^i}-\pi^a \mathcal{L}_T\psi_a+A_t^i G_i+N^aH_a+\bar{\psi}_t S+NH\right)
\label{C.eq:2.11.1}
\end{align}
with $G_i$, $H_a$ and $H$ the Gauss, vector and Hamilton constraint, respectively, as well as an additional constraint $S$ called the SUSY constraint generating local SUSY transformation on the phase space. The phase space is generated by the gravitational electric field $E_a^i=\sqrt{q}e_a^i$ with $q$ the induced metric on $\Sigma$ and the Asthekar-Barbero connection $\tensor*[^{\beta\!\!}]{A}{_a^i}=\Gamma_a^i+\beta K_a^i$ as well as the pair $(\psi_a,\pi^a)$ consisting of the pullback of the Rarita-Schwinger field $\psi_a$ to $\Sigma$ together with its canonically conjugate momentum $\pi^a$ fixed by the reality condition
\begin{equation}
    \Omega^a:=\pi^a+\kappa^{-1}\tensor{\epsilon}{^{abc}}\bar{\psi}_{b}\gamma_{c}\mathcal{P}_{\beta}=0
    \label{C.eq:2.100}
\end{equation}
The reality condition \eqref{C.eq:2.100} is quite complicated due to its explicit dependence on the spatial metric. Thus, to simplify it, one introduces new canonically conjugate fermionic variables by setting
\cite{Sawaguchi:2001wi}
\begin{align}
\phi_i=\frac{\sqrt[4]{q}}{\sqrt{\kappa}}e^a_i\psi_a\quad\text{and}\quad\pi_{\phi}^i=\frac{\sqrt{\kappa}}{\sqrt[4]{q}}e_a^i\pi^a
\label{C.eq:3.1}
\end{align}
As can be checked directly, this indeed provides a canonical transformation on the phase space by redefining the Asthekar-Barbero connection according to
\begin{align}
\tensor*[^{\beta\!\!}]{A}{_a^i}\rightarrow\tensor*[^{\beta\!\!}]{A}{_a^{\prime i}}=\Gamma_a^i+\beta K'^i_a\quad\text{with}\quad
K'^i_a=K_a^i+\frac{\kappa}{\sqrt{q}}\tensor{\epsilon}{^{ijk}}e_a^l\bar{\phi}_{j}\gamma_{k}\mathcal{P}_{\beta}\phi_l
\label{C.eq:3.3.1}
\end{align}
As a consequence, this yields the new canonically conjugate pairs $(A'^i_a,E^a_i)$ and $(\phi_i,\pi^i_{\phi})$ with the nonvanishing Poisson brackets
\begin{align}
\{\tensor*[^{\beta\!\!}]{A}{_a^{\prime i}}(x),E^b_j(y)\}=\kappa\beta\delta^{(3)}(x,y)\quad\text{and}\quad\{\phi_{i}^{\alpha}(x),\pi_{\phi\,\beta}^j(y)\}=-\delta^j_i\delta^{\alpha}_{\beta}\delta^{(3)}(x,y)
\label{C.eq:3.3.2}
\end{align}
In the new variables, the reality condition (\ref{C.eq:2.100}) takes the simplified form
\begin{align}
\Omega^i:=\pi_{\phi}^i+\tensor{\epsilon}{^{ijk}}\bar{\phi}_{j}\gamma_{k}\mathcal{P}_{\beta}=0
\label{C.eq:3.3.3}
\end{align}
and thus no longer depends on the spatial metric. As shown in \cite{BTTVI,Eder:2020uff}, this allows to canonically quantize the theory adapting standard tools of LQG \cite{ThiemannKinematicalHilbertSpaces}. The constraints of the theory also have to be re-expressed in terms of the new variables. For instance, for the SUSY constraint one finds
\begin{align}
S=&i\tensor{\epsilon}{^{abc}}e_a^i\gamma_i\gamma_{*}D_b^{(\tensor*[^{\beta\!\!}]{A}{^{\prime}})}\left(\frac{1}{\sqrt[4]{q}}e^j_c\phi_j\right)+\frac{1}{\sqrt[4]{q}}\tensor{\epsilon}{^{abc}}e_c^l\mathcal{P}_{\beta}\gamma_k(D_a^{(\tensor*[^{\beta\!\!}]{A}{^{\prime}})}e_b^k)\phi_l\nonumber\\
&+\frac{\kappa}{\sqrt[4]{q}}\tensor{\epsilon}{^{ijk}}\gamma^l\phi_{[l}\left(\bar{\phi}_{i]}\mathcal{P}_{\beta}\gamma_k\phi_j\right)+\frac{\kappa\beta}{2\sqrt[4]{q}}\gamma_0\phi^{i}\left(\bar{\phi}_{j}\gamma_k\mathcal{P}_{\beta}\gamma_i\gamma^{(j}\phi^{k)}\right)\nonumber\\
&+\frac{\kappa}{4\sqrt[4]{q}}\gamma_0\phi^{i}\left(\tensor{\epsilon}{^{jkl}}\bar{\phi}_{j}\gamma_0\mathcal{P}_{\beta}\gamma_k\gamma_{i}\phi_{l}\right)-\frac{\sqrt[4]{q}}{L}\gamma^{0i}\phi_i
\label{C.eq:3.22}
\end{align}
In the canonical formulation of SUGRA, the SUSY constraint $S$ plays a major role. This is due to the fact the Poisson bracket $\{S,S\}$ turns out to be proportional to the Hamiltonian constraint \cite{Sawaguchi:2001wi} (see also \cite{Jacobson:1987cj} in the context of the chiral theory as well as \cite{Eder:2020okh} in the context of symmetry reduced models). Hence, the SUSY constraint is superior to the Hamiltonian constraint in the sense that finding solution of the Hamiltonian constraint operator and, thus, solving the dynamics in the quantum theory amounts to finding solutions of the corresponding SUSY constraint operator. This underlines the important role of the SUSY constraint in canonical SUGRA and the necessity to study its implementation in the quantum theory. Besides this, however, the SUSY constraint operator might also be of pure academic interest due to its strong connection to the Hamiltonian constraint via the constraint algebra which may also fix some of the quantization ambiguities. In fact, this has been shown explicitly in the context of symmetry reduced models \cite{Eder:2020okh}.

For the rest of this section, let us briefly outline the quantization of the theory as well as the implementation of the SUSY constraint operator. For more details, we refer the interested reader to \cite{Eder:2020uff}. As shown in \cite{BTTVI}, in order to solve the reality condition \eqref{C.eq:3.3.3}, it is worthwhile to enlarge the phase space by decomposing the fermionic variables in the form $\phi_i=\rho_i+\frac{1}{3}\gamma_i\sigma$ with $\sigma:=\gamma^i\phi_i$ and $\rho_i$ the trace-free part of $\phi_i$ defined via the secondary constraint $\Xi:=\gamma^i\rho_i=0$. Using this constraint together with the reality condition \eqref{C.eq:3.3.3}, one finds that $(\rho_i,\sigma)$ satisfy the following nonvanishing Dirac brackets \cite{Eder:2020uff}
\begin{equation}
    \{\rho^{\alpha}_i(x),\rho_{j}^{\beta}(y)\}_{\mathrm{DB}}=i\mathbf{P}^{\alpha\beta}_{ij}\delta^{(3)}(x,y)\quad\text{and}\quad\{\sigma^{\alpha}(x),\sigma^{\beta}(y)\}_{\mathrm{DB}}=-\frac{9i}{2}\delta_{ij}\delta^{\alpha\beta}\delta^{(3)}(x,y)
\end{equation}
with
\begin{equation}
    \mathbf{P}^{\alpha\beta}_{ij}:=\delta_{ij}\delta^{\alpha\beta}-\frac{1}{3}\tensor{(\gamma_i\gamma_j)}{^{\alpha\beta}}
    \label{C.eq:4.9}
\end{equation}
the projection onto the subspace of trace-free Rarita-Schwinger
fields. Due to the fact that $\mathbf{P}^{\alpha\beta}_{ij}$ defines a projection operator allows to quantize the fermionic fields by adapting standard tools of LQG as developed in \cite{ThiemannKinematicalHilbertSpaces}. Hence, by applying a specific regularization procedure, to $\rho_i$ and $\sigma$ one associates operators $\widehat{\theta}^{(\rho)}_i(x)\equiv\mathbf{P}_{ij}\widehat{\theta}^{(\rho)}_j(x)$ and $\widehat{\theta}^{(\sigma)}(x)$ localized at each point $x\in\Sigma$ acting on a Hilbert space $\mathfrak{H}_{x}$ of certain Grassmann-valued functions on a supermanifold. These operators satisfy the following the anticommutation relations
\begin{equation}
    [\widehat{\theta}^{(\rho)}_i(x),\widehat{\theta}^{(\rho)}_j(y)]=\hbar\mathbf{P}_{ij}\delta_{x,y}\quad\text{and}\quad[\widehat{\theta}^{(\sigma)}(x),\widehat{\theta}^{(\sigma)}(y)]=\frac{9\hbar}{2}\mathds{1}\delta_{x,y}
\end{equation}
The quantized Rarita-Schwinger field on $\mathfrak{H}_x$ is then defined via
\begin{equation}
    \widehat{\theta}_i(x):=\widehat{\theta}^{(\rho)}_i(x)+\frac{1}{3}\gamma_i\widehat{\theta}^{(\sigma)}(x)
\end{equation}
The full Hilbert space $\mathfrak{H}_f$ of the quantized fermionic degrees of freedom is defined as the inductive limit of the tensor product Hilbert spaces $\mathfrak{H}_{\{x_1,\ldots,x_k\}}:=\bigotimes_{i=1}^k\mathfrak{H}_{x_i}$ associated to a finite collection of points $\{x_1,\ldots,x_k\}$ in $\Sigma$.

The bosonic degrees of freedom given by the transformed Asthekar-Barbero connection \eqref{C.eq:3.3.1} and its canonically conjugate momentum $E^a_i$ are quantized in the standard way by studying the associated holonomies and electric fluxes leading to a Hilbert space $\mathfrak{H}_{\mathrm{grav}}$ generated by $\mathrm{SU}(2)$ spin network states. As the total Hilbert space $\mathfrak{H}^{\mathrm{LQSG}}$ of loop quantum supergravity (LQSG) we thus have
\begin{equation}
    \mathfrak{H}^{\mathrm{LQSG}}=\mathfrak{H}_{\mathrm{grav}}\otimes\mathfrak{H}_{f}
\end{equation}
On this Hilbert space, we finally want to implement the SUSY constraint
\begin{equation}
    S[\eta]:=\int_{\Sigma}\mathrm{d}^3x\,\bar{\eta}S=\sum_{i=1}^6S^{(i)}[\eta]
\end{equation}
which, according to \eqref{C.eq:3.22}, splits into 6 subcomponents $S^{(i)}[\eta]$, $i=1,\ldots,6$. As argued in \cite{Eder:2020uff}, the form \eqref{C.eq:3.22} of the SUSY constraint is preferred due to its simplicity and the fact that it does not explicitly involve the extrinsic curvature which, by Thiemann's prescription \cite{ThiemannQSD1}, is related to the Hamiltonian constraint via the Poisson bracket which should be avoided. However, this then requires an alternative quantization procedure. It has been shown in \cite{Eder:2020uff} that all the components $S^{(1)}[\eta]$ can indeed be implemented rigorously and consistently on $\mathfrak{H}^{\mathrm{LQSG}}$. In what follows, let us only comment briefly on the first one $S^{(1)}[\eta]$. By choosing a triangulation of $\Sigma$ similar as in \cite{ThiemannQSD1}, one finds that the quantum analog of $S^{(1)}[\eta]$ schematically takes the form (for more details, we refer to \cite{Eder:2020uff})
\begin{align}
    \widehat{S}^{(1)}[\eta]:=-\frac{1}{3\hbar^2\kappa^2}\sum_{v\in V(\gamma)}
    &\frac{8}{E(v)}
    \bar{\eta}(x_i)i\epsilon^{IJK}\gamma_j\gamma_*[\widehat{\mathscr{X}}_K(s_J(\Delta))-\widehat{\mathscr{X}}_K(x)]\mathrm{tr}(\tau_j h_{s_K(\Delta)}[h_{s_I(\Delta)}^{-1},\widehat{V}_v])
    \label{C.eq:3.1.14}
\end{align}
with
\begin{equation}
    \widehat{\mathscr{X}}_K(s_J(\Delta)):=\mathrm{tr}(\tau_k h_{s'_{K}(\Delta)}[h_{s'_K(\Delta)}^{-1},\sqrt{\widehat{V}}_{s_J(\Delta)}])H_{s_J(\Delta)}\widehat{\theta}_k(s_J(\Delta))
    \label{C.eq:3.1.15}
\end{equation}
As explained in \cite{Eder:2020uff}, in order for the difference $\widehat{\mathscr{X}}_K(s_J(\Delta))-\widehat{\mathscr{X}}_K(x)$ to be nontrivial and thus to correspond to a true quantum analog of the regularized covariant derivative, for $\widehat{V}_v$ one needs to take the Rovelli-Smolin variant of the volume operator $\widehat{V}_v\equiv\widehat{V}^{\mathrm{RS}}_v$ \cite{Rovelli:1994ge,DePietri:1996tvo,Lewandowski:1996gk} as this operator does not act trivially on coplanar vertices in general. However, by choosing different forms of the SUSY constraint, also other quantizations have been proposed in \cite{Eder:2020uff} using the Ashtekar-Lewandowski volume operator.

It follows from \eqref{C.eq:3.1.14} and \eqref{C.eq:3.1.15} that the SUSY constraint operator, while acting on a spin network state, creates new vertices coupled to fermions. Hence, physical solutions of this operator necessary need to contain both bosonic and fermionic degrees of freedom which is consistent with SUSY. The specific form of such solutions, however, has to be investigated in much more detail in the future. Also, it has be checked whether the constraint algebra is anomaly-free and how this relates to the standard quantization scheme of the Hamiltonian constraint.

\subsection{The chiral theory and manifest SUSY}
As seen in the previous section, the SUSY constraint operator in LQG takes a very complicated form. In fact, it acquires an even more complex structure than the ordinary Hamiltonian constraint operator in the bosonic theory. This makes the finding of its solutions very difficult and thus the study of the dynamics in the quantum theory. Moreover, this makes this theory almost inaccessible for direct physical applications such as in the context of black holes or cosmology.

Interestingly, this situation changes drastically in the case of the chiral theory. In fact, the SUSY constraint turns out to be much simpler in this case. In particular, in \cite{Fulop:1993wi} it has been observed that there exists some remnant supersymmetry in the constraint algebra. Based on this observation, in \cite{Gambini:1995db,Ling:1999gn,Ling:2000ss} a quantization has been proposed that allows to keep supersymmetry more manifest in the theory. So far, the construction of the quantum theory remained rather formal. Moreover, the (geometric) origin of the hidden supersymmetry in the constraint algebra as well as possible generalizations to include real values of the Barbero-Immirzi parameter, extended supersymmetry as well as the boundary theory remained unclear. In the following, we review the results of \cite{Eder:2022gge,Eder:2022mft,Eder:2022eqo,Eder:2021rgt} where all these questions are addressed in a unified geometric way, starting from the Holst-MacDowell-Mansouri action of AdS SUGRA \eqref{D.eq:6.4}. We then outline the construction of the quantum theory including SUSY spin nets and area operator and finally discuss applications to black holes and cosmology.

\subsubsection{The chiral Holst-MacDowell-Mansouri action}
In general, the operator $\mathbf{P}_{\mathcal{\beta}}$ contained in the definition \eqref{eq:deformedIP} of the $\beta$-deformed inner product does not commute with local $\mathrm{OSp}(\mathcal{N}|4)$ gauge transformations as, for instance, it acts trivially on gauge transformations corresponding to infinitesimal spacetime translations. Thus, the Holst-MacDowell-Mansouri action, in general, is not manifestly invariant under $\mathrm{OSp}(\mathcal{N}|4)$ gauge transformations. Since, by \eqref{A.eq:I.2.3.22}, the anticommutator of two fermionic generators of opposite chirality generates infinitesimal spacetime translations, it follows that the action can be manifestly invariant under some remnant supersymmetry iff $\mathbf{P}_{\beta}$ singles out fermionic fields of specific chirality. Obviously, this happens only when the Barbero-Immirzi parameter takes the values $\beta=\pm i$. In fact, in this case, it follows that the operator $\mathbf{P}_{-i}$ decomposes in the form $\mathbf{P}_{-i}=\tilde{\mathbf{P}}_{-i}\circ\mathbf{P}^{\mathfrak{osp}(\mathcal{N}|2)}$ with $\mathbf{P}^{\mathfrak{osp}(\mathcal{N}|2)}:\,\mathfrak{osp}(\mathcal{N}|4)\rightarrow\mathfrak{osp}(\mathcal{N}|2)_{\mathbb{C}}$ the projection operator onto the (complexified) chiral sub superalgebra $\mathfrak{osp}(\mathcal{N}|2)_{\mathbb{C}}$ of $\mathfrak{osp}(\mathcal{N}|4)$. Applying this projection operator onto the super Cartan connection \eqref{D.eq:5.1.0} this yields a $\mathfrak{osp}(\mathcal{N}|2)_{\mathbb{C}}$ gauge field
\begin{equation}
\mathcal{A}^{+}:=\mathbf{P}^{\mathfrak{osp}(\mathcal{N}|2)}\mathcal{A}=A^{+ i}T_i^{+}+\psi^A_r Q_A^r+\frac{1}{2}\hat{A}_{rs}T^{rs}
\label{D.eq:SuperAshtekarGeneral}
\end{equation}
containing the bosonic self-dual Ashtekar connection $A^{+i}=\Gamma^i-i K^i$ for which reason it is also referred to as the super Asthekar connection. Using super Asthekar connection, it follows that the Holst-MacDowell-Mansouri action in the chiral limit takes the intriguing form
\begin{align}
S^{\beta=-i}_{\text{H-MM}}(\mathcal{A})=\frac{i}{\kappa}\int_{M}{\braket{F(\mathcal{A^+})\wedge\mathcal{E}}+\frac{1}{4L^2}\braket{\mathcal{E}\wedge\mathcal{E}}}+S_{\text{bdy}}(\mathcal{A}^+)
\label{D.eq:7.8}
\end{align}
with $\mathcal{E}$ the super electric field canonically conjugate to $\mathcal{A}^+$ and transforming under the Adjoint representation of $\mathrm{OSp}(\mathcal{N}|2)_{\mathbb{C}}$. The boundary action $S_{\text{bdy}}(\mathcal{A}^+)$ of the theory is given by
\begin{equation}
S_{\text{bdy}}(\mathcal{A}^+)\equiv S_{\mathrm{CS}}(\mathcal{A}^+)=\frac{k}{4\pi}\int_{H}{\braket{\mathcal{A}^+\wedge\mathrm{d}\mathcal{A}^++\frac{1}{3}\mathcal{A}^+\wedge[\mathcal{A}^+\wedge\mathcal{A}^+]}}
\label{D.eq:7.12}
\end{equation}
with $H:=\partial M$ and thus, in particular, corresponds to the action of a $\mathrm{OSp}(\mathcal{N}|2)_{\mathbb{C}}$ super Chern-Simons theory with (complex) Chern-Simons level $k=i4\pi L^2/\kappa=-i12\pi/\kappa\Lambda$. According to the discussion at the end of section \ref{sec:HolstMM}, it follows that this boundary action as arising from \eqref{D.eq:6.4} in the chiral limit is unique if one imposes supersymmetry invariance at the boundary (see also \cite{Eder:2022mft,Eder:2021rgt}). Moreover, condition \eqref{eq.BoundaryCondition} implies that the chiral projection $\mathbf{P}_{-i}F(\mathcal{A})$ of the curvature has to vanish at the boundary which turns out to be equivalent to the boundary condition
\begin{equation}
    \underset{\raisebox{1pt}{$\Longleftarrow$}}{F(\mathcal{A}^{+})}=-\frac{2\pi i}{\kappa k}\underset{\raisebox{1pt}{$\Leftarrow$}}{\mathcal{E}}
		\label{eq:BoundaryCondition-ChiralTheory}
\end{equation}
where the arrow denotes pullback to the boundary $H$. Condition \eqref{eq:BoundaryCondition-ChiralTheory} imposes the coupling between bulk and boundary degrees of freedom and, as will be discussed in section \ref{sec:BHs} below, plays a major role in studying boundaries in the quantum theory such as the context of quantum supersymmetric black holes.
\subsubsection{The quantum theory and SUSY spin nets}\label{sec:BulkTheory}
Having discussed the geometric structure the Holst-MacDowell-Mansouri action in the chrial limit, following \cite{Eder:2022eqo,Eder:2022gge,Eder:2022mft}, let us next turn to the quantum theory. By varying the bulk action in \eqref{D.eq:7.8}, one obtains
\begin{align}
\delta S_{\text{bulk}}(\mathcal{A})&=\frac{i}{\kappa}\int_{M}{\braket{D^{(\mathcal{A}^+)}\delta\mathcal{A^+}\wedge\mathcal{E}}}=:\mathrm{d}\Theta+\frac{i}{\kappa}\int_{M}\braket{\delta\mathcal{A}^+\wedge D^{(\mathcal{A}^+)}\mathcal{E}}
\label{D.eq:7.9}
\end{align}
with pre-symplectic potential $\Theta(\delta)$ which induces the bulk pre-symplectic structure
\begin{align}
    \Omega_{\text{bulk}}(\delta_1,\delta_2)&=\frac{2i}{\kappa}\int_{\Sigma}\braket{\delta_{[1}\mathcal{A}^+\wedge\delta_{2]}\mathcal{E}}=\frac{i}{\kappa}\int_{\Sigma}{\mathrm{d}^3x\,\left(\delta_1\mathcal{A}^{+\underline{A}}_a\delta_2\mathcal{E}^a_{\underline{A}}-\delta_2\mathcal{A}^{+\underline{A}}_a\delta_1\mathcal{E}^a_{\underline{A}}\right)}
    \label{DH:eq:0.1}
\end{align}
Here
\begin{equation}
\mathcal{E}^a_{\underline{A}}:=\frac{1}{2}\epsilon^{abc}\mathscr{T}_{\underline{B}\underline{A}}\mathcal{E}^{\underline{B}}_{bc}
\label{DH.eq:1.1}
\end{equation}
where we have chosen a homogeneous basis $(T_{\underline{A}})_{\underline{A}}$ of $\mathfrak{osp}(\mathcal{N}|2)_{\mathbb{C}}$ and set $\mathscr{T}_{\underline{A}\underline{B}}:=\braket{T_{\underline{A}},T_{\underline{B}}}$. Hence, it follows that $(\mathcal{A}^{+\underline{A}}_a,\mathcal{E}^a_{\underline{A}})$ generate a graded symplectic phase space of canonical chiral SUGRA and satisfy the graded Poisson bracket
\begin{equation}
    \{\mathcal{E}^a_{\underline{A}}(x),\mathcal{A}_b^{+\underline{B}}(y)\}=i\kappa\delta^a_b\delta^{\underline{B}}_{\underline{A}}\delta^{(3)}(x,y)
    \label{DH:eq:0.3}
\end{equation}
Moreover, from \eqref{D.eq:7.9} we can immediately read off the bulk super Gauss constraint
\begin{equation}
    \mathscr{G}[\alpha]:=\frac{i}{\kappa}\int_{\Sigma}\braket{\alpha\wedge D^{(\mathcal{A}^+)}\mathcal{E}}=-\frac{i}{\kappa}\int_{\Sigma}\braket{\mathcal{E}\wedge D^{(\mathcal{A}^+)}\alpha}+\frac{i}{\kappa}\int_{\Delta}\braket{\mathcal{E},\alpha}
    \label{GaussN=2+bdy}
\end{equation}
where $\alpha$ is some $\mathfrak{osp}(\mathcal{N}|2)_{\mathbb{C}}$-valued smearing function and $\Delta:=\Sigma\cap\partial M$. This constraint satisfies $\{\mathscr{G}[\alpha],\mathscr{G}[\beta]\}=\mathscr{G}[[\alpha,\beta]]$ and therefore generates local $\mathrm{OSp}(\mathcal{N}|2)_{\mathbb{C}}$-gauge transformations on the phase space. Hence, it follows that the canonical chiral theory has an intriguing structure which turns out to be in complete analogy the bosonic self-dual theory. This suggests to quantize the bulk theory adapting and generalizing tools of standard LQG. In the following, let us briefly outline the construction of the quantum theory following \cite{Eder:2022eqo,Eder:2022gge}. The mathematical details underlying this construction can be found in \cite{Eder:2022mft,Eder:2021ans}.

In standard LQG, the bulk Hilbert space $\mathfrak{H}^{\mathrm{bulk}}_{\gamma}$ associated to a graph $\gamma$ embedded in the spatial slices $\Sigma$ of the spacetime manifold $M=\mathbb{R}\times\Sigma$ is obtained by considering spin network states, a class of states invariant under local gauge transformations. They are constructed via contraction of matrix coefficients of irreducible representations of the underlying gauge group. For this it is crucial that the representations under consideration form a tensor category. Finite-dimensional irreducible representations of the orthosymplectic series $\mathrm{OSp}(\mathcal{N}|2)$ for $\mathcal{N}=1,2$ have been intensively studied (see e.g. \cite{Scheunert:1976wj,Scheunert:1976wi,Minnaert:1990sz,Berezin:1981jz}. For the case $\mathcal{N}=1$, these representations form a subcategory closed under the tensor product. The corresponding spin network states have been studied for instance in \cite{Gambini:1995db,Ling:1999gn}.
For the case $\mathcal{N}=2$ the subclass of \emph{typical representations} form such a category (see \cite{Scheunert:1976wj}).

We now describe the construction of the super spin network states for a suitable subclass $\EuScript{P}_{\mathrm{adm}}$ of irreducible representations (finite- or infinite-dimensional, possibly including those constructed in \cite{Eder:2022gge}) of $\mathrm{OSp}(\mathcal{N}|2)$ with $\mathcal{N}=1,2$. For any subset $\Vec{\pi}:=\{\pi_{e}\}_{e\in E(\gamma)}\subset\EuScript{P}_{\mathrm{adm}}$, we define the cylindrical function $T_{\gamma,\Vec{\pi},\Vec{m},\Vec{n}}$ via
\begin{equation}
    T_{\gamma,\Vec{\pi},\Vec{m},\Vec{n}}[\mathcal{A}^+]:=\prod_{e\in E(\gamma)}\tensor{\pi_e(h_e[\mathcal{A}^+])}{^{m_e}_{n_e}}
    \label{DH.eq:3.18}
\end{equation}
where, for any edge $e\in E(\gamma)$, $h_e[\mathcal{A}^+]$ denotes the super holonomy (parallel transport) of the connection $\mathcal{A}^+$ along $e$ given by\footnote{It is interesting to note that the proper definition of the holonomy of a super connection requires the consideration of a more general class of supermanifolds called relative supermanifolds (see \cite{Eder:2021ans}) that depend on a certain parametrization. In this way, it follows that holonomies take values in a generalized Lie supergroup.} (see \cite{Eder:2022mft,Eder:2021ans} for more details)
\begin{equation}
    h_e[\mathcal{A}^+]:=\mathcal{P}\mathrm{exp}\left(\int_{e}{\mathrm{Ad}_{h_{e}[A]}\psi}\right)\cdot h_e[A]
\end{equation}
where we split $\mathcal{A}^+$ in the form $\mathcal{A}^+=A+\psi$ with and $A$ and $\psi$ the underlying bosonic and fermionic subcomponent, respectively, with $A:=A^++\hat{A}$ consisting of the self-dual Ashtekar connection $A^+$ and the $\mathfrak{so}(\mathcal{N})$-gauge field $\hat{A}$. Furthermore, $\tensor{(\pi_e)}{^{m_e}_{n_e}}$ denote certain matrix coefficients of the representation $\pi_e\in\EuScript{P}_{\mathrm{adm}}$. In order to get a gauge invariant state, at each vertex $v\in V(\gamma)$ of the graph $\gamma$, we have to contract \eqref{DH.eq:3.18} with an intertwiner $I_v$ projecting onto the trivial representation at any vertex. As a result, the so-constructed state transforms trivially under local gauge transformations, i.e., it is annihilated by the quantum analog of the super Gauss constraint \eqref{GaussN=2+bdy}, and thus indeed forms a gauge-invariant state which we call a (gauge-invariant) super spin network state. We take these states as a basis of the state space of the bulk theory. We assume that an inner product can be found that turns this space into a super Hilbert space $\mathfrak{H}^{\mathrm{bulk}}_{\gamma}$.

In this context, it is important to emphasize that one still needs to implement reality conditions as, a priori, one is dealing with complex variables. Following similar arguments as in the bosonic theory \cite{Frodden:2012dq,BenAchour:2014erw}, it seems highly suggestive that a solution of the reality conditions is given by selecting a specific subclass of representations of $\mathrm{OSp}(\mathcal{N}|2)_{\mathbb{C}}$ with respect to which geometric operators have particular nice and consistent properties. We will describe these kind of representations in the following section. In section \ref{sec:BHs}, we will demonstrate that these representation indeed lead to consistent semi-classical results in the context of black hole computation.

\subsubsection{The super area operator}


On the space of super spin networks, one can introduce a gauge-invariant operator in analogy to the area operator in standard LQG (see also Sec. \ref{sec:area higher D} in the context of higher dimensions). More precisely, since the super electric field $\mathcal{E}$ defines a $\mathfrak{osp}(\mathcal{N}|2)_{\mathbb{C}}$-valued 2-form, for any oriented (semianalytic) surface $S$ embedded in $\Sigma$, one can define the graded or super area quantity $\mathrm{gAr}(S)$ via
\begin{equation}
\mathrm{gAr}(S):=\sqrt{2}\int_{S}{\|\mathcal{E}\|}
\label{Chapter5:area1}
\end{equation}
where, in analogy to \cite{Eder:2018uzm,Corichi:2000dm,Ashtekar:2000hw}, the norm $\|\mathcal{E}\|$ on $S$ is defined via $\|\mathcal{E}\|:=\sqrt{\braket{\mathcal{E}_S,\mathcal{E}_S}}$ with $\mathcal{E}_S$ the unique $\mathfrak{osp}(\mathcal{N}|2)_{\mathbb{C}}$-valued function on $S$ such that $\iota^*_S\mathcal{E}=\mathcal{E}_S\,\mathrm{vol}_S$ with $\mathrm{vol}_S$ the volume-form on $S$. For  the special case $\mathcal{N}=1$, the expression \eqref{Chapter5:area1} coincides with the super area as considered in \cite{Ling:1999gn}. Here, the prefactor $\sqrt{2}$ has been chosen such that in the case of vanishing fermionic degrees of freedom, the super area reduces to the standard area of $S$ in ordinary Riemannian geometry.

By definition, the quantity \eqref{Chapter5:area1} solely depends on the super electric field which defines a phase space variable. Thus, we can implement it in the quantum theory by performing a similar regularization procedure in the bosonic theory (see \cite{Eder:2022mft,Eder:2022gge} for more details). As a result, for $\mathcal{N}=1$, it follows for instance in the case that the surface $S$ intersects the graph $\gamma$ of a (gauge-invariant) super spin network state $T_{\gamma,\Vec{\pi},\Vec{m},\Vec{n}}$ labeled by super spin quantum numbers $j\in\mathbb{C}$ corresponding to the principal series representations of $\mathrm{OSp}(1|2)$ as constructed in \cite{Eder:2022gge} in a single divalent vertex $v\in V(\gamma)$ that the action of super area operator is given by
\begin{equation}
\widehat{\mathrm{gAr}}(S) T_{\gamma,\Vec{\pi},\Vec{m},\Vec{n}}=-8\pi il_p^2\sqrt{j\left(j+\frac{1}{2}\right)}T_{\gamma,\Vec{\pi},\Vec{m},\Vec{n}}
\label{eq:3.3.49}
\end{equation}
with $j\in\mathbb{C}$ the superspin quantum number labeling the edge $e\in E(\gamma)$ intersecting the surface $S$. For $j\in\frac{N_0}{2}$, this coincides with the result of \cite{Ling:1999gn}.

According to \eqref{eq:3.3.49} the super area operator has complex eigenvalues which seems to be physically inconsistent. In fact, so far, we have not implemented the reality conditions in the quantum theory. Similar as in the context of the bosonic theory \cite{Frodden:2012dq,BenAchour:2014erw}, one may then argue that solving the reality conditions amounts to select a specific subclass of representations with respect to which the spectrum of the super area operator becomes purely real. Interestingly, as shown in \cite{Eder:2022gge}, it turns out that the principal series of $\mathrm{OSp}(1|2)_{\mathbb{C}}$ indeed contains a subclass of irreducible representations that lead to a real spectrum of the super area operator. To be more precise, if one considers the series of representations labeled by superspin quantum numbers of the form
\begin{align}
    j=-\frac{1}{4}+is\text{ with }s\in\mathbb{R}
    \label{eq:4.2.21}
\end{align}
one finds that the action of the super area operator on these super soin network states takes the form
\begin{equation}
\widehat{\mathrm{gAr}}(S) T_{\gamma,\Vec{\pi},\Vec{m},\Vec{n}}=8\pi l_p^2\sqrt{s^2+\frac{1}{16}}\,T_{\gamma,\Vec{\pi},\Vec{m},\Vec{n}}
\label{eq:4.2.23}
\end{equation}
and therefore it follows that super spin network states whose edges are labeled by $j$ satisfying \eqref{eq:4.2.21} are indeed eigenstates of the super area operator with real eigenvalues. Interestingly, this is in complete analogy to the bosonic theory \cite{Frodden:2012dq}.
\subsection{Application: Black hole entropy}\label{sec:BHs}
In this final section, we briefly want to review the results of the articles \cite{Eder:2022eqo,Eder:2022gge,Eder:2022mft} studying physical applications of the formalism as outlined above. More precisely, in \cite{Eder:2022eqo,Eder:2022gge} inner boundaries in chiral loop quantum supergravity have been considered. In particular, the possibility to associate an entropy to such boundaries has been discussed by studying microstates generated by super spin network states piercing the boundary. In the macroscopic limit, the resulting entropy formula remarkably turns out to be consistent with semi-classical computations. Let us also note that in \cite{Eder:2020okh} applications of the formalism in the context of minisuperspace models have been discussed. There, a complete solution of the reality conditions as well as a consistent implementation of the constraint algebra has been achieved. However, to keep it short, we will not review these results in this section and refer the interested reader to the original literature.

\begin{figure}
     \centering
     \includegraphics[height=4cm]{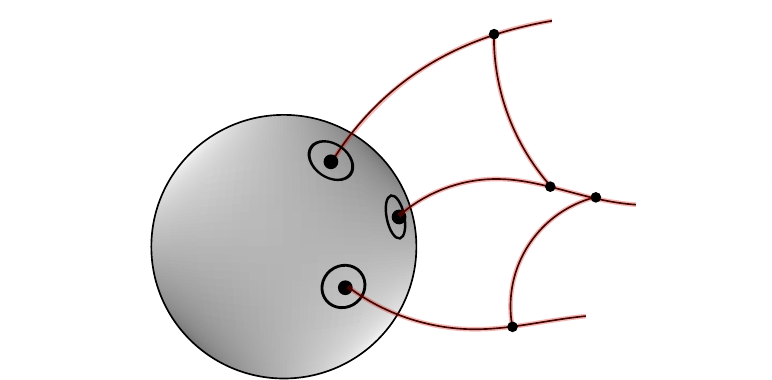}
     \caption{Visualization of a supersymmetric black hole in chiral LQSG. The super spin network states induce non-trivial Chern-Simons degrees of freedom (black circles) at the intersection points (punctures) with the boundary which can account for black hole entropy (source \cite{Eder:2022mft}).}
     \label{fig:BH}
 \end{figure}

As described in Section \ref{sec:BulkTheory} above, the quantum excitations of the bulk degrees of freedom are represented by super spin network states associated to the gauge supergroup $\mathrm{OSp}(\mathcal{N}|2)_{\mathbb{C}}$. On the other hand, according to Equations \eqref{D.eq:7.8} and \eqref{D.eq:7.12}, the unique boundary theory is described in terms of a $\mathrm{OSp}(\mathcal{N}|2)_{\mathbb{C}}$ super Chern-Simons theory. Hence, for a given finite graph $\gamma$ embedded in $\Sigma$, one may define the Hilbert space $\mathfrak{H}_{\text{full},\gamma}$ w.r.t. $\gamma$ of the full theory as the tensor product
\begin{equation}
    \mathfrak{H}^{\text{full}}_{\gamma}=\mathfrak{H}^{\mathrm{bulk}}_{\gamma}\otimes\mathfrak{H}^{\text{bdy}}_{\gamma}
    \label{Gauss:fullTheory1}
\end{equation}
with $\mathfrak{H}^{\mathrm{bulk}}_{\gamma}$ the Hilbert space of the quantized bulk degrees of freedom as constructed in Section \ref{sec:BulkTheory} and $\mathfrak{H}^{\text{bdy}}_{\gamma}$ the Hilbert space corresponding to the quantized super Chern-Simons theory on the boundary.\\
On this Hilbert space, one needs to implement the boundary condition \eqref{eq:BoundaryCondition-ChiralTheory}. To this end, at each puncture $p\in\EuScript{P}_{\gamma}:=\gamma\cap\Delta$, one chosses a disk $D_{\epsilon}(p)$ on $\Delta$ around $p$ with radius $\epsilon>0$ and set
\begin{equation}
    \mathcal{E}[\alpha](p):=\lim_{\epsilon\rightarrow 0}\int\limits_{D_{\epsilon}(p)}\braket{\alpha,\mathcal{E}},\quad F[\alpha](p):=\lim_{\epsilon\rightarrow 0}\int\limits_{D_{\epsilon}(p)}\braket{\alpha,F(\mathcal{A}^+)}
    \label{Gauss:fullTheory2}
\end{equation}
By definition, these quantities (or suitable functions thereof) can be promoted to well-defined operators in the quantum theory. Thus, \eqref{eq:BoundaryCondition-ChiralTheory} yields the additional constraint equation
\begin{equation}
\mathds{1}\otimes\widehat{F}_{\underline{A}}(p)=-\frac{2\pi i}{\kappa k}\widehat{\mathcal{E}}_{\underline{A}}(p)\otimes\mathds{1}
\label{Gauss:fullTheory4}
\end{equation}
at each puncture $p\in\EuScript{P}_{\gamma}$, in analogy to the bosonic theory \cite{Engle:2009vc,Kaul:1998xv}. The quantized super electric flux $\widehat{\mathcal{E}}_{\underline{A}}(p)$ acts in terms of right- resp. left-invariant vector fields (see \cite{Eder:2022gge,Eder:2022mft}). Hence, from \eqref{Gauss:fullTheory4}, one concludes that the Hilbert space of the quantized boundary degrees of freedom corresponds to the Hilbert space of a quantized $\mathrm{OSp}(\mathcal{N}|2)_{\mathbb{C}}$ super Chern-Simons theory on $\Delta$ with punctures $\EuScript{P}_{\gamma}$ and complex Chern-Simons level $k$ (see figure \ref{fig:BH}).
According to Boltzmann, this suggests to associate an entropy to the boundary by defining it as the logarithm of the the number of Chern-Simons degrees of freedom generated by the super spin network edges piercing the boundary. Unfortunately, the (super) Chern-Simons theory with complex and non-compact gauge group is not well-known. Moreover, it is not clear how to deal with the fact that the Chern-Simons level is purely imaginary. Interestingly, similar issues also seem to arise in the context of boundary theories in string theory \cite{Mikhaylov:2014aoa}.

One may therefore adapt the strategy of \cite{BenAchour:2014erw} in the context of the purely bosonic theory to the supersymmetric setting by studying a specific compact real form of $\mathrm{OSp}(\mathcal{N}|2)_{\mathbb{C}}$ and then performing an analytic continuation to the corresponding complex Lie supergroup. More precisely, restricting to the minimal supersymmetric case $\mathcal{N}=1$ in what follows, one considers the Chern-Simons theory with compact gauge supergroup $\mathrm{UOSp}(1|2)$ and integer Chern-Simons level $k=-12\pi/\kappa\Lambda_{\mathrm{cos}}$ and punctures labeled by finite-dimensional irreducible representations $\Vec{j}$ of $\mathrm{UOSp}(1|2)$ with $j\in\frac{\mathbb{N}_0}{2}$. One then computes the number $\mathcal{N}_k(\Vec{j})$ of Chern-Simons degrees of freedom given by the dimension of the superconformal blocks. Finally, one performs an analytic continuation by replacing $j\rightarrow j=-\frac{1}{4}+is$ for some $s\in\mathbb{R}$ for each $j\in\Vec{j}$ as well as $k\rightarrow ik$ in $\mathcal{N}_k(\Vec{j})$. In order to simplify the discussion, let us assume that the boundary $H$ is topologically of the form $\mathbb{R}\times\mathbb{S}^2$, that is, the $2$-dimensional slices $\Delta_t$ are topologically equivalent to $2$-spheres. Furthermore, we consider the limit $k\rightarrow\infty$ corresponding to a vanishing cosmological constant $\Lambda\rightarrow 0$. Under these assumptions, it follows that the number of microstates $\mathcal{N}_{\infty}(\Vec{j})$  is given by the number of $\mathrm{UOSp}(1|2)$ gauge-invariant states, i.e., it can be identified with the number of trivial subrepresentations contained in the tensor product representation $\bigotimes_j \pi_j$. As shown in \cite{Eder:2022gge}, $\mathcal{N}_{\infty}$ can be expressed by the following integral formula
\begin{align}
    \mathcal{N}_{\infty}(\{n_l,j_l\})=\frac{1}{2\pi}\int_0^{\pi}\mathrm{d}\theta\,\sin^2(2\theta)\left[4-n+\sum_{i=1}^p n_i d_{j_l}\frac{\tan(d_{j_i}\theta)}{\tan\theta}\right]\prod_{l=1}^p\left(\frac{\cos(d_{j_l}\theta)}{\cos\theta}\right)^{n_l}
    \label{eq:VerlindeFormula}
\end{align}
where $\Vec{j}$ has been subdivided into $p\leq n:=|\EuScript{P}_{\gamma}|$ subfamilies $(n_l,j_l)$, $l=1,\ldots,p$, corresponding to $0<n_l\leq n$ punctures labeled by $j_l\in\Vec{j}$. Moreover, $d_j:=4j+1$ denotes the dimension of the finite-dimensional representation of $\mathrm{UOSp}(1|2)$ labeled by $j\in\frac{\mathbb{N}_0}{2}$. By performing an analytic continuation of \eqref{eq:VerlindeFormula} following the algorithm as described above, one finds that the analytically continued state sum formula can be written as
\begin{align}
  \mathcal{I}_{\infty}=\frac{2}{\pi}\int_{\mathcal{C}}\mathrm{d}z&\,\mu(z)\left[1-\frac{n}{4}-\sum_{i=1}^p n_is_i\frac{\tan(s_i z)}{\tanh z}\right]\exp\left(\sum_{l=1}^p n_l\ln\left(\frac{\cos(s_l z)}{\cosh z}\right)\right)
    \label{eq:5.3.1}
\end{align}
where $\mu(z):=i\sinh^2(2z)$ and $\mathcal{C}$ is a contour from 0 to $i\pi$. The integral formula \eqref{eq:5.3.1} can be evaluated approximately in the macroscopic limit corresponding to a large number of punctures $n_l\rightarrow\infty$ as well as large colors $s_l\rightarrow\infty$ $\forall l=1,\ldots,p$. In this limit and in the case of $n$ indistinguishable punctures, one then finds that the entropy defined via $S:=\ln(|\mathcal{I}_{\infty}|/n!)$ is given by
\begin{equation}
    S=\ln\left(\frac{|\mathcal{I}_{\infty}|}{n!}\right)=\frac{a_H}{4l_p^2}+\mathcal{O}(\sqrt{a_H})
    \label{eq:5.3.8}
\end{equation}
with $a_H:=8\pi\sum_{l=1}^p n_ls_l$ the area of the boundary as measured w.r.t. the super area operator \eqref{eq:3.3.49}. Hence, in the macroscopic limit, it follows that the entropy associated to the boundary is indeed proportional to the area of the boundary in highest order with the correct prefactor of $1/4$ as predicted by the semi-classical computations of Bekenstein and Hawking \cite{Bekenstein:1973ur,Hawking:1975vcx}. Furthermore, the entropy acquires lower order quantum corrections that can also be computed explicitly in the macroscopic limit (see \cite{Eder:2022gge} for more details).

It is important to emphasize that eq. \eqref{eq:5.3.8} follows here directly from the analytically continued state sum formula \eqref{eq:5.3.1}. In particular, we did not have to make any choices or fix the Barbero-Immirzi parameter to specific values. Moreover, this confirms the results of \cite{Frodden:2012dq,BenAchour:2014erw} in the bosonic theory and supports the hypothesis that, in the context of complex variables, the entropy can be derived via an analytic continuation starting from a compact real form of the complex gauge group. In the future, it would be very desirable to generalize these results to include supegravity theories with extended supersymmetry. In this respect, the pure $D=4$, $\mathcal{N}=4$ case would be of particular interest as it has been intensively studied in the string theory literature \cite{Strominger:1996sh}.

In any case, this example demonstrates that the techniques as discussed in this review can indeed be used for concrete physical applications leading to results that are consistent with semi-classical computations and which may provide a starting point for linking and testing ideas from LQG and string theory.

\section{Modified gravity}

Historically, GR serves as the
simplest relativistic theory of gravity with correct Newtonian limit. It is worth pursuing all alternatives, which
provide a high chance to uncover new physics. Recently, Loop quantum gravity(LQG) has been generalized to the metric $f(\R)$
theories \cite{Zh11,Zh11b}, Brans-Dicke theory \cite{ZM12a} and
scalar-tensor theories (both in four-dimensional case \cite{ZM11c} and higher-dimensional case \cite{Han14}) and lowest-order projectable Ho\v{r}ava-Lifshitz gravity \cite{Ma20}. The fact that this background-independent quantization method can be successfully extended to those modified
theories of gravity rely on the key observation that these gravity theories
can also be reformulated into the connection dynamical formalism with
a compact structure group. The purpose of this paper is to review how to get the connection dynamics of these modified gravity theories and how to quantize these theories by the nonperturbative loop quantization procedure. We will use the Scalar-Tensor Theories(STT) as the representative theories to illustrate our methodology. We only focus on the canonical quantization methods and omit the cosmological applications of loop quantum modified gravity which can be found in Refs. \cite{Zhang13,Haro14,Gupt16}.

\subsection{General scheme\label{section2}}
In this section, we will first outline the general scheme of loop
quantization for metric modified gravity theories \cite{Ma12a}. Here we are
mainly focusing on 4-dimensional metric theories of gravity and assume the modified gravity theory which is under consideration has a well-defined geometrical dynamics description. In this sense, the Hamiltonian formalism with 3-metric $q_{ab}$ as one of
the configuration variables is required. Moreover, the constraint algebra should be first-class (or after solving some second-class
constraints). Without loss of generality, we suppose the
classical phase space of this theory consists of conjugate pairs
$(q_{ab},p^{ab} )$ and $(\phi_A,\pi^A)$ with $\phi_A$ being a
scalar, vector, tensor or spinor field. Then our quantization scheme has the following recipe.

1, To obtain the corresponding connection dynamical formalism of modified gravity theories, we first introduce a quantity
$\kt_{ab}$ via \ba \kt_{ab} =
\frac{16\pi G}{\sqrt{h}}\left(p_{ab}-\frac12pq_{ab}\right).\ea Then
we enlarge the phase space to obtain the triad formulation
as \ba(q_{ab},p^{ab} ) \Rightarrow  (E^a_i \equiv\sqrt{q}q_{ab}
e^b_i , \kt^i_a\equiv\kt_{ab}e^b_i ).\ea Now we make a canonical
transformation to connection formulation as: \ba (E^a_j ,
\kt^j_a)\Rightarrow (E^a_j ,A^j_a\equiv\Gamma^j_a
+\gamma\kt^j_a),\ea and to recover the symmetric property of $p_{ab}$ we
have to impose the condition $ \kt_{a[i}E^a_{j]} =0$. This turns out to be the desired Gaussian
constraint, $\mathcal {D}_aE^a_i\equiv \partial_aE^a_i +
\epsilon_{ijk}A^j_aE^a_k=0$ which justified the internal structure group we adopted. By using these new variables, we can write all
the constraints in terms of the new variables straightforwardly.

2, For loop quantization, we represent the fields $(\phi_A,\pi^A)$ via
polymer-like representation, together with the LQG representation
for the holonomy-flux algebra for the gravitational part. Then the resulted kinematical Hilbert space
is $\hil_\kin := \hil^\grav_\kin \otimes\hil^\phi_\kin$.
All the basic operators and geometrical operators could be promoted as the well-defined operators in this Hilbert space. As in the standard LQG, the Gaussian and
diffeomorphism constraints can be solved. Then we would get the
gauge invariant $\hil_G$ and diffeomorphism covariant Hilbert spaceas $\hil_{Diff}$ respectively. In
order to implement quantum dynamics, the Hamiltonian constraint
operator can be constructed at least in $\hil_G$ or even at $\hil_{Diff}$, although it
usually could not be well defined in $\hil_{Diff}$.

Due to the extreme complication, the quantum dynamics of LQG is still an open issue. Thus in the following sections, we will take STT as an example to carry out the steps 1 and 2 in the above scheme.

\subsection{Hamiltonian dynamics \label{section3}}

The most general action of STT reads \ba
S(g)=\frac{1}{16\pi G}\int_\Sigma
d^4x\sqrt{-g}\left[\phi\R-\frac{\omega(\phi)}{\phi}(\partial_\mu\phi)\partial^\mu\phi-2V(\phi)\right]\label{action}
\ea where $\R$ denotes the scalar curvature of
spacetime metric $g_{\mu\nu}$, and the coupling parameter $\omega(\phi)$
and potential $V(\phi)$ being the coupling parameter and potential respectively. By
doing 3+1 decomposition of the spacetime, the four-dimensional
scalar curvature has the following decomposition \ba \mathcal
{R}=K_{ab}K^{ab}-K^2+R+\frac{2}{\sqrt{-g}}\partial_\mu(\sqrt{-g}n^\mu
K)-\frac{2}{N\sqrt{q}}\partial_a
(\sqrt{q}q^{ab}\partial_bN)\label{03} \ea where $K_{ab}$ is the
extrinsic curvature of the 3-dimensional spatial metric $q_{ab}$, $K\equiv
K_{ab}q^{ab}$, $R$ denotes the scalar curvature on the spatial surface $\Sigma$, $n^\mu$ is the unit normal of $\Sigma$
and $N$ represents the lapse function. By Legendre transformation, the
momenta conjugate to the dynamical variables $q_{ab}$ and $\phi$ are
defined respectively as
\ba p^{ab}&=&\frac{\partial\mathcal
{L}}{\partial\dot{q}_{ab}}=\frac{\sqrt{q}}{16\pi G}[\phi(K^{ab}-Kq^{ab})-\frac{h^{ab}}{N}(\dot{\phi}-N^c\partial_c\phi)], \label{04}\\
\pi&=&\frac{\partial\mathcal
{L}}{\partial\dot{\phi}}=-\frac{\sqrt{q}}{8\pi G}(K-\frac{\omega(\phi)}{N\phi}(\dot{\phi}-N^c\partial_c\phi)),\label{pi}
\ea with $N^c$ being the shift vector. Combining of the trace of Eq.
(\ref{04}) with Eq. (\ref{pi}) gives us\ba
(3+2\omega(\phi))(\dot{\phi}-N^a\partial_a\phi)=\frac{16\pi G
N}{\sqrt{h}}(\phi\pi-p).\label{Sconstraint} \ea From Eq. (\ref{Sconstraint}), it is easy to see that one extra constraint
$S=p-\phi\pi=0$ emerges when $\omega(\phi)=-\frac32$. Hence we
divided the theories into two sector with
$\omega(\phi)\neq-\frac32$ and $\omega(\phi)=-\frac32$.

\subsubsection{Sector of $\omega(\phi)\neq -3/2$ }

In the $\omega(\phi)\neq -3/2$ sector, the Hamiltonian of STT can
be derived as a liner combination of constraints similar to GR as
\ba H_{total}=\int_\Sigma d^3x(N^aV_a+NH),\label{htotal} \ea
with
the smeared diffeomorphism and Hamiltonian constraints being
\ba V(\overrightarrow{N})&=&\int_\Sigma d^3xN^aV_a =\int_\Sigma
d^3xN^a\left(-2D^b(p_{ab})+\pi\partial_a\phi\right),\label{dc}\\
H(N)&=&\int_\Sigma d^3xNH \ea
with \ba
H=\frac{16\pi G}{\sqrt{q}}\left(\frac{p_{ab}p^{ab}-\frac12p^2}{\phi}+\frac{(p-\phi\pi)^2}{2\phi(3+2\omega)}\right)+\frac{\sqrt{q}}{16\pi G}\left(-\phi
R+\frac{\omega(\phi)}{\phi}(D_a\phi)
D^a\phi+2D_aD^a\phi+2V(\phi)\right).\label{hc}\nn\\
\ea
By using the symplectic structure
\ba
\{q_{ab}(x),p^{cd}(y)\}&=&\delta^{(c}_a\delta^{d)}_b\delta^3(x,y),\nn\\
\{\phi(x),\pi(y)\}&=&\delta^3(x,y), \label{poission}\ea
straightforward calculation shows that the constraints (\ref{dc})
and (\ref{hc}) comprise a first-class system similar to GR as:
\ba \{V(\overrightarrow{N}),V(\overrightarrow{N}^\prime)\}&=& V([\overrightarrow{N},\overrightarrow{N}^\prime]), \nn\\
\{H(M),V(\overrightarrow{N})\}&=&- H(\mathcal
{L}_{\overrightarrow{N}}M), \nn\\
\{H(N),H(M)\}&=& V(ND^aM-MD^aN). \ea

To further obtain the connection dynamical formalism of the STT, following
the general scheme, we introduce
\ba\tilde{K}^{ab}=\phi
K^{ab}+\frac{q^{ab}}{2N}(\dot{\phi}-N^c\partial_c\phi)=\phi
K^{ab}+\frac{q^{ab}}{(3+2\omega)\sqrt{q}}(\phi\pi-p).\ea
Then we can construct new canonical pairs
$(E^a_i\equiv\sqrt{h}e^a_i, \tilde{K}_a^i\equiv\kt_{ab}e^b_i)$, where
$e^a_i$ is the 3-triad such that $q_{ab}e^a_ie^b_j=\delta_{ij}$.
Now
the symplectic structure (\ref{poission}) implies the
following non-zero Poisson brackets: \ba
\{\tilde{K}^j_a(x),E_k^b(y)\}=8\pi G\delta^b_a\delta^j_k\delta(x,y).
\ea Other Possion brackets are all vanished. Note that the symmetric property $\tilde{K}^{ab}=\tilde{K}^{ba}$, we need to impose an additional constraint: \be
G_{jk}\equiv\tilde{K}_{a[j}E^a_{k]}=0. \label{gaussian}\ee
Now we
can make a second canonical transformation via defining: \be
A^i_a=\Gamma^i_a+\gamma\tilde{K}^i_a, \label{newvaribles}\ee
where
$\Gamma^i_a$ is the spin connection, and
$\gamma$ is a nonzero real number. It is easy to check that $A^j_a$ goes back to the Ashtekar-Barbero connection
\cite{BarberoRealAshtekarVariables} for $\phi=1$. The non-vanish Poisson bracket
among the new variables reads: \ba
\{A^j_a(x),E_k^b(y)\}&=&8\pi G\gamma\delta^b_a\delta^j_k\delta(x,y).\ea
By combining Eq.(\ref{gaussian}) with
the compatibility condition $\nabla_aE^a_i=\partial_aE^a_i+\epsilon_{ijk}\Gamma^j_aE^{ak}=0$ give us
the standard Gaussian constraint
\ba \mathcal
{G}_i=\mathscr{D}_aE^a_i\equiv\partial_aE^a_i+\epsilon_{ijk}A^j_aE^{ak}=0,
\label{GC}\ea
which justifies $A^i_a$ is indeed an $su(2)$-connection. The
diffeomorphism and Hamiltonian constraints in terms of new variables up to Gaussian constraint read respectively as \ba
V_a &=&\frac1{8\pi G\gamma} F^i_{ab}E^b_i+\pi\partial_a\phi,
\label{diff}\ea
\ba H
&=&\frac{\phi}{16\pi G}\left[F^j_{ab}-(\gamma^2+\frac{1}{\phi^2})\varepsilon_{jmn}\tilde{K}^m_a\tilde{K}^n_b\right]
\frac{\varepsilon_{jkl}
E^a_kE^b_l}{\sqrt{q}}\nn\\
&+&\frac{1}{(3+2\omega(\phi))}\left(\frac{(\tilde{K}^i_aE^a_i)^2}{8\pi G\phi\sqrt{q}}+
2\frac{(\tilde{K}^i_aE^a_i)\pi}{\sqrt{q}}+\frac{8\pi G\pi^2\phi}{\sqrt{q}}\right) \nn\\
&+&\frac{1}{8\pi G}\left[\frac{\omega(\phi)}{2\phi}\sqrt{q}(D_a\phi)
D^a\phi+\sqrt{q}D_aD^a\phi+\sqrt{q}V(\phi)\right],\label{hamilton}
\ea where
$F^i_{ab}\equiv2\partial_{[a}A^i_{b]}+\epsilon^i_{\ kl}A_a^kA_b^l$ represents
the curvature of connection $A_a^i$. The total Hamiltonian thus can be expressed as
a linear combination
\ba H_{total}=\ints\Lambda^i\mathcal
{G}_i+N^aV_a+NH.\ea
It is not hard to check that the smeared Gaussian
constraint $\mathcal {G}(\Lambda):=\int_\Sigma
d^3x\Lambda^i(x)\mathcal {G}_i(x)$ generates $SU(2)$ internal gauge
transformations, while the smeared diffeomorphism constraint
$\mathcal {V}(\overrightarrow{N}):=\int_\Sigma
d^3xN^a(V_a-A_a^i\mathcal {G}_i)$ generates spatial diffeomorphism
transformations. Together with the smeared
Hamiltonian constraint $H(N)=\int_\Sigma d^3xNH$,
the constraints algebra has the following form: \ba \{\mathcal
{G}(\Lambda),\mathcal {G}(\Lambda^\prime)\}&=&8\pi G\mathcal
{G}([\Lambda,\Lambda^\prime]),\label{eqsA} \\
\{\mathcal
{G}(\Lambda),\mathcal{V}(\overrightarrow{N})\}&=&-\mathcal{G}(\mathcal
{L}_{\overrightarrow{N}}\Lambda,),\\
\{\mathcal {G}(\Lambda),H(N)\}&=&0,\\
\{\mathcal {V}(\overrightarrow{N}),\mathcal
{V}(\overrightarrow{N}^\prime)\}&=&\mathcal
{V}([\overrightarrow{N},\overrightarrow{N}^\prime]), \\
\{\mathcal {V}(\overrightarrow{N}),H(M)\}&=& H(\mathcal
{L}_{\overrightarrow{N}}M),\label{eqsE}\\
\{H(N),H(M)\}&=&\mathcal {V}(ND^aM-MD^aN)\nn\\
&+&\mathcal
{G}\left((N\partial_aM-M\partial_aN)q^{ab}A_b\right)\nn\\
&-&\frac{[E^aD_aN,E^bD_bM]^i}{8\pi G q}\mathcal {G}_i\nn\\
&-&\gamma^2\frac{[E^aD_a(\phi N),E^bD_b(\phi M)]^i}{8\pi G q}\mathcal {G}_i.\label{eqsb}\ea The STT of gravity in the sector
$\omega(\phi)\neq -3/2$ now have already been cast into the
$su(2)$-connection dynamical formalism which lays the foundation for further loop quantization.

\subsubsection{Sector of $\omega(\phi)= -3/2$ }

Now we turn to the  sector of $\omega(\phi)= -3/2$. Eq. (\ref{Sconstraint})
suggests that there exists an extra primary constraint $S=0$, which we call
``conformal" constraint, the name conformal will became clear in the later on. Hence, the total Hamiltonian in this sector can be
written as a liner combination \ba
H_{total}=\int_\Sigma d^3x(N^aV_a+NH+\lambda S),\label{htotal1} \ea
where the smeared diffeomorphism constraint $V(\overrightarrow{N})$ is the same
as (\ref{dc}), while the Hamiltonian and conformal constraints read
respectively \ba
H(N)&=&\int_\Sigma d^3xNH \nn\\
&=&\int_\Sigma
d^3xN\left[\frac{16\pi G}{\sqrt{q}}\left(\frac{p_{ab}p^{ab}-\frac12p^2}{\phi}\right)
+\frac{\sqrt{q}}{16\pi G}(-\phi R-\frac{3}{2\phi}(D_a\phi)
D^a\phi+2D_aD^a\phi+2V(\phi))\right],\label{hc1}\nn\\
S(\lambda)&=&\int_\Sigma d^3x\lambda S=\int_\Sigma
d^3x\lambda(p-\phi\pi).\label{sc}\ea
With the help of symplectic
structure (\ref{poission}), straightforward calculations show that \ba
\{H(M),V(\overrightarrow{N})\}&=&- H(\mathcal
{L}_{\overrightarrow{N}}M),\quad
\{S(\lambda),V(\overrightarrow{N})\}=- S(\mathcal
{L}_{\overrightarrow{N}}\lambda),\label{VHS}\\
\{H(N),H(M)\}&=& V(ND^aM-MD^aN)+ S(\frac{D_a\phi}{\phi}(ND^aM-MD^aN)),\label{HH}\\
\{S(\lambda),H(M)\}&=& H(\frac{\lambda M}{2})+\ints
N\lambda\sqrt{q}(-2V(\phi)+\phi V'(\phi)).\label{Sc} \ea
Eq. (\ref{Sc}) implies that, a
secondary constraint must to impose to insure to preserve the constraints $S$ and $H$ as
\ba -2V(\phi)+\phi V'(\phi)=0.
\label{equationofV}\ea It is clear that this constraint is
second-class, and hence one has to solve it. For the
vacuum case where the solutions of Eq. (\ref{equationofV}) are strictly constrained into either
$V(\phi)=0$ or $V(\phi)=C\phi^2$, with $C$ being constant. For these two theories, the action
(\ref{action}) is invariant under the following conformal
transformation:
\ba \tilde{g}_{\mu\nu}\rightarrow e^\lambda g_{\mu\nu},\quad
\tilde{\phi}\rightarrow e^{-\lambda}\phi.\label{conformalt} \ea
Thus,
besides diffeomorphism invariance, those theories has an extra conformal symmetry. The resulted Hamiltonian formalism of the theory comprise of a set of the
constraints $(V,H,S)$. Moreover, these constraints form a first-class system, in particular, the
transformations on the phase space which generated by the conformal
constraint are as follows
\ba &&\{q_{ab},S(\lambda)\}=\lambda h_{ab},\quad
\{P^{ab},S(\lambda)\}=-\lambda P^{ab}, \\
&&\{\phi,S(\lambda)\}=-\lambda \phi,\quad \{\pi,S(\lambda)\}=\lambda
\pi. \ea
Due to this extra conformal symmetry (\ref{sc}), the resulted physical
degrees of freedom of this special kind of STT are equal to those of
GR.

By the canonical transformations and shift to the new variables (\ref{newvaribles}), the resulted connection-dynamics of STT can be obtained and the total Hamiltonian can be expressed as
\ba H_{total}=\ints\Lambda^i\mathcal
{G}_i+N^aV_a+NH+\lambda S,\ea
where the Gaussian and diffeomorphism
constraints are the same as in sector $\omega(\phi)\neq-3/2$,
while the Hamiltonian and the conformal constraints read
respectively
\ba  H
&=&\frac{\phi}{16\pi G}\left[F^j_{ab}-(\gamma^2+\frac{1}{\phi^2})\varepsilon_{jmn}\tilde{K}^m_a\tilde{K}^n_b\right]
\frac{\varepsilon_{jkl}
E^a_kE^b_l}{\sqrt{q}}\nn\\
&+&\frac{1}{8\pi G}\Big[-\frac{3}{4\phi}\sqrt{q}(D_a\phi)
D^a\phi+\sqrt{q}D_aD^a\phi+\sqrt{q}V(\phi)\Big],\label{hamilton1}\\
S&=&\frac{1}{8\pi G}\tilde{K}^i_aE^a_i-\pi\phi
\label{conformalc}.\ea
These constraints again forms a first class system \cite{ZM11c}.

\subsection{Loop quantum kinematics and dynamics of scalar-tensor theories\label{section4}}

With the connection dynamics in hand,
the nonperturbative loop quantization procedure now is ready to
extend to the STT. The kinematical structure of STT keeps the same form as
that of standard LQG coupled with a polymer scalar field \cite{Zh11,Zh11b} which means the kinematical
Hilbert space STT is also a direct product of the Hilbert
space of the gravitational part and that of the polymer scalar field,
$\hil_\kin:=\hil^\grav_\kin\otimes \hil^\sca_\kin$. This Hilbert space $\hil_\kin$ admits an orthonormal
basis which is so-called spin-scalar-network
basis over some graph $\alpha\cup X\subset\Sigma$ as \ba
T_{\alpha,X}(A,\phi)\equiv T_{\alpha}(A)\otimes T_{X}(\phi),\ea
where $\alpha$ and $X$ consist of finite number of curves and points in $\Sigma$
respectively. The fundamental quantum operators of STT are the holonomy operator $h_e(A)=\mathcal {P}\exp\int_eA_a$ defined by a
connection along edges $e\subset\Sigma$ and densitized triads operator
$E(S,f):=\int_S\epsilon_{abc}E^a_if^i$ which smeared over
2-surfaces, wile for the scalar field part, the basic operator are the point holonomies
$U_\lambda=\exp(i\lambda\phi(x))$, and momenta
$\pi(R):=\int_R d^3x\pi(x)$ smeared on 3-dimensional regions. As the characteristically feature of LQG, the spatial geometric operators are the same as those in standard LQG. Moreover, it is also natural
to promote the Gaussian constraint $\mathcal {G}(\Lambda)$ as a
well-defined operators\cite{ThiemannModernCanonicalQuantum} by the standard LQG way. The kernel of Gaussian constraint operator gives the
internal gauge invariant Hilbert space $\mathcal {H}_G$. Similarly to LQG, the diffeomorphism constraint can be solved by the group averaging technique and the desired diffeomorphism covariant Hilbert spapce $\mathcal {H}_{Diff}$ will be obtained \cite{ThiemannModernCanonicalQuantum,ZM11c}. The remaining task then is to implement the quantum Hamiltonian operators on this gauge invariant and diffeomorphism covariant Hilbert space.

\subsubsection{Sector of $\omega(\phi)\neq -3/2$ }

In order to implement the
Hamiltonian constraint (\ref{hamilton}) at the quantum level. Let us first write Eq.
(\ref{hamilton}) as $H(N)=\sum^8_{i=1}H_i$ in regular order. The complete treatment of the Hamiltonian constraint can be found in \cite{ZM11c}. here we just take the term $H_6=\int_\Sigma
d^3xN\frac{\omega(\phi)}{2\phi}\sqrt{h}(D_a\phi) D^a\phi $ as a example to give the complete process. This
term is somehow like the kinetic term of a Klein-Gordon field. To begin with, we first introduce the well-defined
operators $\hat{\phi},\widehat{\phi^{-1}}$ \cite{Zh11b}. This enables us to quantize that function $\omega(\phi)$ in the way of Taylor expansion. Moreover, By the point-spliting regularization techniques as in
Refs. \cite{Ma06,Zh11b}, we first triangulate the spatial surface $\Sigma$ in adaptation to
some graph $\alpha$ underling a cylindrical function in $\hil_\kin$. Moreover,
we reexpress connections by basic variables such as holonomies or triads. The action of corresponding regulated
operator on a basis vector $T_{\alpha,X}$ reads \ba \hat{H}^\varepsilon_6\cdot T_{\alpha,X}
&=&\lim_{\epsilon\rightarrow 0}\frac{2^{17}N(v)\hat{\omega}(\phi)
}{3^6\gamma^4(i\lambda_0)^2(i\hbar)^48\pi G}\hat{\phi}^{-1}(v)
\chi_\epsilon(v-v')
\nn\\
&\times&\sum_{v\in\alpha(v)}\frac{1}{E(v)}\sum_{v(\Delta)=v}\epsilon(s_L
s_M s_N)\epsilon^{LMN}\hat{U}^{-1}_{\lambda_0}(\phi(s_{s_L(\Delta_{v})}))\nn\\
&\times&
[\hat{U}_{\lambda_0}(\phi(t_{s_L(\Delta_{v})}))-\hat{U}_{\lambda_0}(\phi(s_{s_L(\Delta_{v})}))]\nn\\
&\times&\Tr(\tau_i\hat{h}_{s_M(\Delta_{v})}[\hat{h}^{-1}_{s_M(\Delta_{v})},(\hat{V}_{U^\epsilon_{v}})^{3/4}]
\hat{h}_{s_N(\Delta_{v})}[\hat{h}^{-1}_{s_N(\Delta_{v})},(\hat{V}_{U^\epsilon_{v}})^{3/4}]) \nn\\
&\times&\sum_{v'\in\alpha(v)}\frac{1}{E(v')}\sum_{v(\Delta')=v'}\epsilon(s_I
s_J s_K)\epsilon^{IJK}\hat{U}^{-1}_{\lambda_0}(\phi(s_{s_I(\Delta_{v'})}))\nn\\
&\times&
[\hat{U}_{\lambda_0}(\phi(t_{s_I(\Delta_{v'})}))-\hat{U}_{\lambda_0}(\phi(s_{s_I(\Delta_{v'})}))]\nn\\
&\times&\Tr(\tau_i\hat{h}_{s_J(\Delta_{v'})}[\hat{h}^{-1}_{s_J(\Delta_{v'})},(\hat{V}_{U^\epsilon_{v'}})^{3/4}]
\hat{h}_{s_K(\Delta_{v'})}[\hat{h}^{-1}_{s_K(\Delta_{v'})},(\hat{V}_{U^\epsilon_{v'}})^{3/4}])\cdot
T_{\alpha,X}. \label{H6}\ea
The detailed meaning
of notations in Eq.(\ref{H6}) can be found in \cite{Zh11b}. It is obvious that the action of
$\hat{H}^\varepsilon_6$ on $ T_{\alpha,X}$ will change the graph. Some vertices at $t(s_I(v))=\varepsilon$ for
edges $e_I(t)$ starting from each high-valent vertex of $\alpha$ are added. As
a consequence, when $\varepsilon\rightarrow 0$, the family of operators $\hat{H}^\varepsilon_6(N)$ is not weakly convergent. We must use the so-called uniform
Rovelli-Smolin topology induced by diffeomorphism-invariant states
$\Phi_{Diff}$ to define the limit operator as:
\ba \Phi_{Diff}(\hat{H}_6\cdot
T_{\alpha,X})=\lim_{\varepsilon\rightarrow
0}(\Phi_{Diff}|\hat{H}^\varepsilon_{6}|T_{\alpha,X}\rangle. \ea
The limit is independent of $\varepsilon$ which means regulator $\varepsilon$ can be removed and the action of the limit operator reads
\ba \hat{H}_6\cdot T_{\alpha,X} &=&\sum_{v\in
V(\alpha)}\frac{2^{17}N(v)\hat{\omega}(\phi) }{3^6\gamma^4(i\lambda_0)^2(i\hbar)^4E^2(v)8\pi G }\hat{\phi}^{-1}(v)\nn\\
&\times&\sum_{v(\Delta)=v(\Delta')=v}\epsilon(s_L
s_M s_N)\epsilon^{LMN}\hat{U}^{-1}_{\lambda_0}(\phi(s_{s_L(\Delta)}))\nn\\
&\times&
[\hat{U}_{\lambda_0}(\phi(t_{s_L(\Delta)}))-\hat{U}_{\lambda_0}(\phi(s_{s_L(\Delta)}))]\nn\\
&\times&\Tr(\tau_i\hat{h}_{s_M(\Delta)}[\hat{h}^{-1}_{s_M(\Delta)},(\hat{V}_v)^{3/4}]
\hat{h}_{s_N(\Delta)}[\hat{h}^{-1}_{s_N(\Delta)},(\hat{V}_v)^{3/4}]) \nn\\
&\times&\epsilon(s_I
s_J s_K)\epsilon^{IJK}\hat{U}^{-1}_{\lambda_0}(\phi(s_{s_I(\Delta')}))\nn\\
&\times&
[\hat{U}_{\lambda_0}(\phi(t_{s_I(\Delta')}))-\hat{U}_{\lambda_0}(\phi(s_{s_I(\Delta')}))]\nn\\
&\times&\Tr(\tau_i\hat{h}_{s_J(\Delta')}[\hat{h}^{-1}_{s_J(\Delta')},(\hat{V}_{v})^{3/4}]
\hat{h}_{s_K(\Delta')}[\hat{h}^{-1}_{s_K(\Delta')},(\hat{V}_{v})^{3/4}])\cdot
T_{\alpha,X} . \ea The first two terms in the Hamiltonian constraint (\ref{hamilton}) are just the Euclidean and Lorentz term multiply with some function of $\phi$. Since the action of polymer scalar field and gravitational part are separated. So the action of these two terms are rather simple. For rest five terms, the detailed expression can be found in Ref.\cite{ZM11c}
Therefore, the total Hamiltonian constraint in this sector has been
fully quantized and promoted as a well-defined operator
$\hat{H}(N)=\sum_{i=1}^8\hat{H}_i$ on kinematical Hilbert space $\hil_\kin$. The construction of $\hat{H}(N)$ insures its internal gauge invariant and diffeomorphism
covariant. Hence the Hamiltonian operator is well defined at least in the gauge-invariant
Hilbert space $\hil_G$. However, it is still difficult to define
$\hat{H}(N)$ directly on $\hil_{Diff}$. Moreover, the constraint algebra of STT does not form a Lie algebra. This
leads to possible quantum anomaly after quantization.

\subsubsection{Sector of $\omega(\phi)= -3/2$ }

In the special case of $\omega(\phi)= -3/2$, we need to promote the smeared conformal constraint (\ref{conformalc}) to be a well-defined operator on kinematical Hilbert space. Note
that both $\phi$ and $\pi(R)$ for polymer scalar field are already well-defined operators. First we note that the following indentity:
\ba \kt\equiv\int_\Sigma
d^3x\tilde{K}^i_aE^a_i=\gamma^{-\frac32}\{\Euc(1),V\} \ea
where the
Euclidean term $\Euc(1)$ is defined as: \ba
\Euc(1)&=&\frac{1}{16\pi G}\int_\Sigma
d^3xF^j_{ab}\frac{\varepsilon_{jkl} E^a_kE^b_l}{\sqrt{h}}. \ea Both
$\Euc$ and the volume $V$ have been quantized in LQG. Then we can easily
promote $S(\lambda)$ as a well-defined operator on a given basis vector $T_{\alpha,X}\in\hil_\kin$ as \ba
\hat{S}(\lambda)\cdot T_{\alpha,X} &=& \left(\sum_{v\in
V(\alpha)}\frac{\lambda(v)}{\gamma^{3/2}8\pi G(i\hbar)}[\hat{H}^E(1),\hat{V}_v]-\sum_{x\in
X}\lambda(x)\hat{\phi}(x)\hat{\pi}(x)\right)\cdot T_{\alpha,X}.\ea
It is clear that $\hat{S}(\lambda)$ is internal gauge invariant and
diffeomorphism covariant, Moreover, it will also change a given graph. Hence it is already well
defined in gauge invariant Hilbert space $\hil_G$. Other terms in the Hamiltonian constraint operator in this
sector is similar to that in the sector of $\omega(\phi)\neq -3/2$.
The difference is that now $\omega$ takes the particular value of $\omega= -3/2$. We write Eq.
(\ref{hamilton1}) as $H(N)=\sum^5_{i=1}H_i$ in regular order. It is
easy to check that the terms $H_1,H_2,H_4,H_5$ keeps the same form
as those in last subsection, while the remaining term $H_3$ can be quantized as
\ba \hat{H}_3\cdot T_{\alpha,X} &=&-\sum_{v\in
V(\alpha)}\frac{2^{16}N(v)}{3^5\gamma^4(i\lambda_0)^2(i\hbar)^48\pi G E^2(v)}\hat{\phi}^{-1}(v)\nn\\
&\times&\sum_{v(\Delta)=v(\Delta')=v}\epsilon(s_L
s_M s_N)\epsilon^{LMN}\hat{U}^{-1}_{\lambda_0}(\phi(s_{s_L(\Delta)}))\nn\\
&\times&
[\hat{U}_{\lambda_0}(\phi(t_{s_L(\Delta)}))-\hat{U}_{\lambda_0}(\phi(s_{s_L(\Delta)}))]\nn\\
&\times&\Tr(\tau_i\hat{h}_{s_M(\Delta)}[\hat{h}^{-1}_{s_M(\Delta)},(\hat{V}_v)^{3/4}]
\hat{h}_{s_N(\Delta)}[\hat{h}^{-1}_{s_N(\Delta)},(\hat{V}_v)^{3/4}]) \nn\\
&\times&\epsilon(s_I
s_J s_K)\epsilon^{IJK}\hat{U}^{-1}_{\lambda_0}(\phi(s_{s_I(\Delta')}))\nn\\
&\times&
[\hat{U}_{\lambda_0}(\phi(t_{s_I(\Delta')}))-\hat{U}_{\lambda_0}(\phi(s_{s_I(\Delta')}))]\nn\\
&\times&\Tr(\tau_i\hat{h}_{s_J(\Delta')})[\hat{h}^{-1}_{s_J(\Delta')},(\hat{V}_{v})^{3/4}]
\hat{h}_{s_K(\Delta')}[\hat{h}^{-1}_{s_K(\Delta')},(\hat{V}_{v})^{3/4}])\cdot
T_{\alpha,X} . \ea Now the total Hamiltonian constraint operator
$\hat{H}(N)=\sum^5_{i=1}\hat{H}_i$ is also a well-defined operator in kinematical Hilbert space
$\hil_G$.

\subsection{Master constraint and the physical Hilbert space\label{section5}}

The quantum dynamics of LQG is remaining an open question, in order to find a viable way to obtain the possible physical Hilbert space, Thiemann introduces the master constraint programme into LQG \cite{Th06}. One can form the master constraint from an anomalous set of constraints. Hence in this section, we generalize the master constraint for the above quantum STT. Parallel to the above sections, we also discuss master constraint with different $\omega(\phi)$.
In the sector of $\omega(\phi)\neq -3/2$, the
master constraint of the STT is similar to that in GR case \cite{ZM11c}. While, for the case of $\omega(\phi)= -3/2$, due to the existence of the extra conformal constraint. The master constraint for this sector should be generalized to include the conformal constraint as \cite{ZM11c}
\ba \mathcal {M}:=\frac12\int_\Sigma
d^3x\frac{\abs{H(x)}^2+\abs{S(x)}^2}{\sqrt{h}}, \label{mcs1}\ea
where the expressions of Hamiltonian constraint $H(x)$ and the
conformal constraint $S(x)$ are given by Eqs. (\ref{hamilton1}) and
(\ref{conformalc}) respectively. It is easy to see that \ba \mathcal {M}=0
\Leftrightarrow H(N)=0 \quad and \quad S(\lambda)=0 \quad\forall
N(x),\lambda(x). \ea
It is not hard to check that now the Master constraint together with Gauss constraint and diffeomorphism constraint form a Lie algebra. After some suitable regularization methods \cite{ZM11c}. The
quantum master constraint operator $\hat{\mathcal {M}}$ acting on diffeomorphism invariant
states as \ba( \hat{\mathcal {M}}\Phi_{Diff})T_{s,c}=\lim_{\mathcal
{P}\rightarrow\Sigma,\varepsilon,\varepsilon'\rightarrow
0}\Phi_{Diff}[\frac12\sum_{c\in\mathcal
{P}}\left(\hat{H}^\varepsilon_C(\hat{H}^{\varepsilon'}_C)^\dagger+\hat{S}^\varepsilon_C(\hat{S}^{\varepsilon'}_C)^\dagger\right)
T_{s,c} ]. \ea The operator $\hat{\mathcal {M}}$ is
positive and symmetric in $\hil_{Diff}$ and admits
a unique self-adjoint Friedrichs extension\cite{Ma06,Zh11b}. Hence
it is also possible to obtain the physical Hilbert space of the
quantum STT in this special case by the direct integral
decomposition of $\hil_{Diff}$ with respect to the spectrum of
$\hat{\mathcal {M}}$.

\section*{Acknowledgements}

NB was supported by an International Junior Research Group grant of the Elite Network of Bavaria. KE was supported by the German Acadamic Scholarship Foundation (Studienstiftung des deutschen Volkes). XZ was supported by National Natural Science Foundation of China (NSFC) with No.12275087, No.11775082 and ``the Fundamental Research Funds for the Central Universities''.


\bibliographystyle{utphysmendeley}
\bibliography{library}



\end{document}